\documentclass[onecolumn,secnumarabic,amssymb, nobibnotes, nofootinbib,aps, prd,showpacs,showkeys,unsortedaddress,10pt]{revtex4-1}
\pdfoutput=1

\def\bmi {\begin{minipage}}
\def\emi {\end{minipage}}
\usepackage{appendix}
\usepackage{epsfig}
\usepackage{graphicx}
\usepackage{color}
\usepackage{soul}
\usepackage{ulem}
\usepackage{amssymb}
\usepackage{hyperref}
\usepackage{multirow}
\usepackage{float}
\def\vecb{\bf b}
\def\vecK{\bf K}
\def\ee{\end{equation}}
\def\be{\begin{equation}}
\def\eea{\end{eqnarray}}
\def\bea{\begin{eqnarray}}
\def\bal{\begin{align}}
\def\eal{\end{align}}

\def\vecb{{\bf b}}
\def\vecb{\bf b}
\def\vecK{\bf K}
\def\veck{\bf k}

\def\be{\begin{equation}}
\def\ee{\end{equation}}
\def\bea{\begin{eqnarray}}
\def\eea{\end{eqnarray}}

\def\bef{\begin{figure}}
\def\ptmin{p_{tmin}}
\def\rs{\sqrt{s}}

\def\eef{\end{figure}}
\begin{document}
\title{Soft edge of hadron scattering and mini-jet models for  the total   and inelastic $pp$ cross sections at LHC and beyond}
\author{D. A. Fagundes}
\email{dfagundes@ift.unesp.br}
\affiliation{Instituto de F\'isica Te\'orica, UNESP, Rua Dr. Bento T. Ferraz, 271, Bloco II, 01140-070, S\~ao Paulo - SP, Brazil}
\author{A. Grau}
\email{igrau@ugr.es}
\affiliation{Departamento de F\'{\i}sica Te\'orica y del Cosmos, Universidad de Granada, 18071 Granada, Spain}
\author{G. Pancheri}
\email{giulia.pancheri@lnf.infn.it}
\affiliation{INFN Frascati National Laboratories, Via E. Fermi 40, 00444, Italy}
\author{Y. N. Srivastava}
\email{yogendra.srivastava@gmail.com}
\affiliation{Physics Department, University of Perugia, 06123 Perugia, Italy}
\author{O. Shekhovtsova}
\email{olga.shekhovtsova@ifj.edu.pl}
\affiliation{Kharkov Institute of Physics and Technology
61108, Akademicheskaya,1, Kharkov, Ukraine\\
Institute of Nuclear Physics PAN ul. Radzikowskiego 152 31-342 Krakow, Poland}
\begin{abstract}
We show that the onset and rise of QCD mini-jets provide the dynamical mechanism behind the appearance
of a {\it soft edge} in $pp$ collisions around ISR energies and thus such a soft edge is {\it built in} our mini-jet 
model with soft gluon re-summation. Here the model is optimized for LHC at $\sqrt{s} = 7, 8\ TeV$ and predictions
made for higher LHC and cosmic ray energies. Further, we provide a phenomenological picture to discuss
the breakup of the total cross section into its elastic, uncorrelated and correlated inelastic pieces in the
framework of a one-channel  eikonal function.      
\end{abstract}

\vskip 3 cm

\maketitle

\section{Introduction}

This paper 
has three main objectives: (i) to  clarify the role played by QCD mini-jets in the high energy behavior of  
total cross sections 
in reference to recent observation about  
the onset of an energy invariant  {\it soft edge}  \cite{Block:2014lna}  in $pp$ collisions, around CERN intersecting storage ring (ISR) energies; (ii) to provide 
an update of our model predictions at 
$\sqrt{s}=13$  and $14 \ TeV$ (LHC13 and LHC14) 
for proton-proton scattering, both for total \cite{Achilli:2007pn} and inelastic \cite{Achilli:2011sw} cross sections  and
(iii) to apply our mini-jet model to the very high energy behavior that may be of utility for analyses of 
the particle production cross section in the realm of  cosmic rays.

The paper is organized as follows. In Sec.(\ref{soft-edge}), we begin by showing how the notion of mini-jets 
in its simplest formulation when augmented by asymptotic freedom leads easily to the appearance of a 
threshold in the total cross section around $\sqrt{s}\simeq (10\div 20)\ GeV$. Through the use of 
current Parton Density Functions (PDFs), we 
then show that the onset and the rise of the mini-jet cross section is behind the observation of the {\it soft edge}. 

Having thus highlighted the role of  the mini jet phenomenon in 
view of the recent observation of an  {\it  edge-like} structure 
\cite{Block:2014lna,Rosner:2014nka} occurring in hadronic collisions near 
$\sqrt{s}\simeq (10\div 20)\ GeV$, in Sec.(\ref{Soft}), we provide 
a brief description of the model we have proposed quite a long time ago \cite{Grau:1999em}, 
which  
overcomes  the main limitation of the  mini-jet description, namely the violation of the Froissart bound,
through  
soft gluon $k_t$ re-summation.  We take the opportunity to optimize our predictions for both LHC and higher energy 
$pp$ data extracted from cosmic rays,  and include recent results at $\sqrt{s}=7$ and $8\ TeV$. 
We also indicate the uncertainty coming from different sets of PDFs
entering the calculations.   
 
In Sec.(\ref{sec:sigs}), we 
apply our model to 
inelastic scattering at LHC and discuss the role played by mini-jets in non-diffractive scattering. A
breakup of the inelastic cross section in terms of its correlated and uncorrelated parts and the extraction
of the measured  elastic cross section is provided in detail for a one-channel eikonal formulation, alongside with 
 comparison with all available data. 

Our model is not in conflict with the dominant analytic description of total, and elastic cross section   
based on Regge-Pomeron 
models such as for instance \cite{Gotsman:2012rm} or \cite{Khoze:2014nia},
 rather it provides  an intuitive description of hadronic scattering 
in terms of low $p_t$ collisions describable by perturbative QCD and soft gluon emission, for which 
a mixed approach, perturbative and non-perturbative is required. 
 \section{{\it Soft edge}  and mini jets}\label{soft-edge}
These are two basic  {\it facts} about the $pp$ and $p{\bar p}$ total cross section:
\begin{itemize}
\item $\sigma_{total}$ first decreases and then, around $\sqrt{s}\simeq (10\div 20)\ GeV$ rises 
as a function of the squared cm energy $s$. 
\item $\sigma_{total}$  asymptotically cannot rise faster than $\log ^2[s/s_o]$
\end{itemize}
Both phenomena have a physical interpretation in terms of  QCD,  the difficulty lies in a proper implementation 
of the second fact. We shall start, in this section, from  the first {\it fact}, in particular   the rise, which appears 
when gluon-gluon scattering becomes observable. This argument can be made quantitative.

At low CM energy, soft gluon emission accompanying non-perturbative effects decreases the cross section, while there are not enough perturbative gluons for beam-on-beam scattering.
  However, as the energy increases, more energy is available for gluon emission, 
and there is a non-zero probability for    hard gluon-gluon scattering with production of final state partons of 
$p_t\sim 1 \ GeV$.   For such energy   regions, 
 the collision can be described by perturbative QCD through the asymptotic freedom expression for the coupling, $\alpha_{AF}(p^2_t)$, valid when $p_t^2>>\Lambda^2_{QCD}$.  
The emission of hard gluons is a perturbative process, characterized by a $1/x\sim \sqrt{s}/(2p_t)$ parton spectrum.
As the energy  increases further, the cross section 
for collisions resulting with a final state parton with $p_t\gtrsim 1$, will  start rising  because of $1/x$ behavior, entering into a {\it high gluon luminosity} region. 
We see that   the  contribution to the total cross section from these hard collisions can become sizable and  rising with energy 
when  
 \begin{equation}
 1/x =\sqrt{s}/2p_t >>1\ and \  p_t>1\ GeV.
 \end{equation} 
With the condition $p_t>1\ GeV$ for the {\it asymptotic freedom condition} to be  satisfied, and the rise starting when $x<<1$, for instance 
$\simeq 0.1-0.2$, the turning point for the rise to start is for 
\begin{eqnarray}
\sqrt{s} \gtrsim (2 / x) GeV  \\
 \sqrt{s}\sim (10\div 20) \ GeV 
\end{eqnarray}
as the data show. The transition from collisions which do not involve scattering of  perturbative gluons, 
to a region dominated by mini-jets, is what 
is seen in the total cross section data, and is discussed, in a somewhat different language,  
in 
\cite{Block:2014lna}.  

The above argument can be made quantitative by calculating 
the bulk of  perturbative gluon-gluon collisions, i.e.   the mini jet cross section.The mini-jet cross section is given by
\begin{equation}
\sigma^{AB}_{\rm jet} (s;\ptmin) = \int_{\ptmin}^{\rs/2} d p_t \int_{4
p_t^2/s}^1 d x_1 \int_{4 p_t^2/(x_1 s)}^1 d x_2 \sum_{i,j,k,l}
f_{i|A}(x_1,p_t^2) f_{j|B}(x_2, p_t^2)~~
 \frac { d \hat{\sigma}_{ij}^{ kl}(\hat{s})} {d p_t}.
\label{sigjet}
\end{equation}
where  $f_{i|A}(x_1,p_t^2)$ are the PDFs with  $i, \ j, \ k, \ l$ to  denote 
the partons and $x_1,x_2$ the
fractions of the parent particle momentum carried by the parton.
$\sqrt{\hat{s}} = \sqrt{x_1 x_2 s}$,   $\hat{ \sigma}$ are the
center of mass energy of the two parton system and the hard parton
scattering cross--section respectively. Following the argument given above, 
this expression sums only collisions with outgoing partons of momentum  with $p_t>p_{tmin}$, 
where $p_{tmin}$ is defined as the region of validity of perturbative QCD, i.e. the coupling is 
given by the asymptotic freedom expression  for running  $\alpha_F$. 
\begin{figure}[h]
\resizebox{1\textwidth}{!}{
\includegraphics{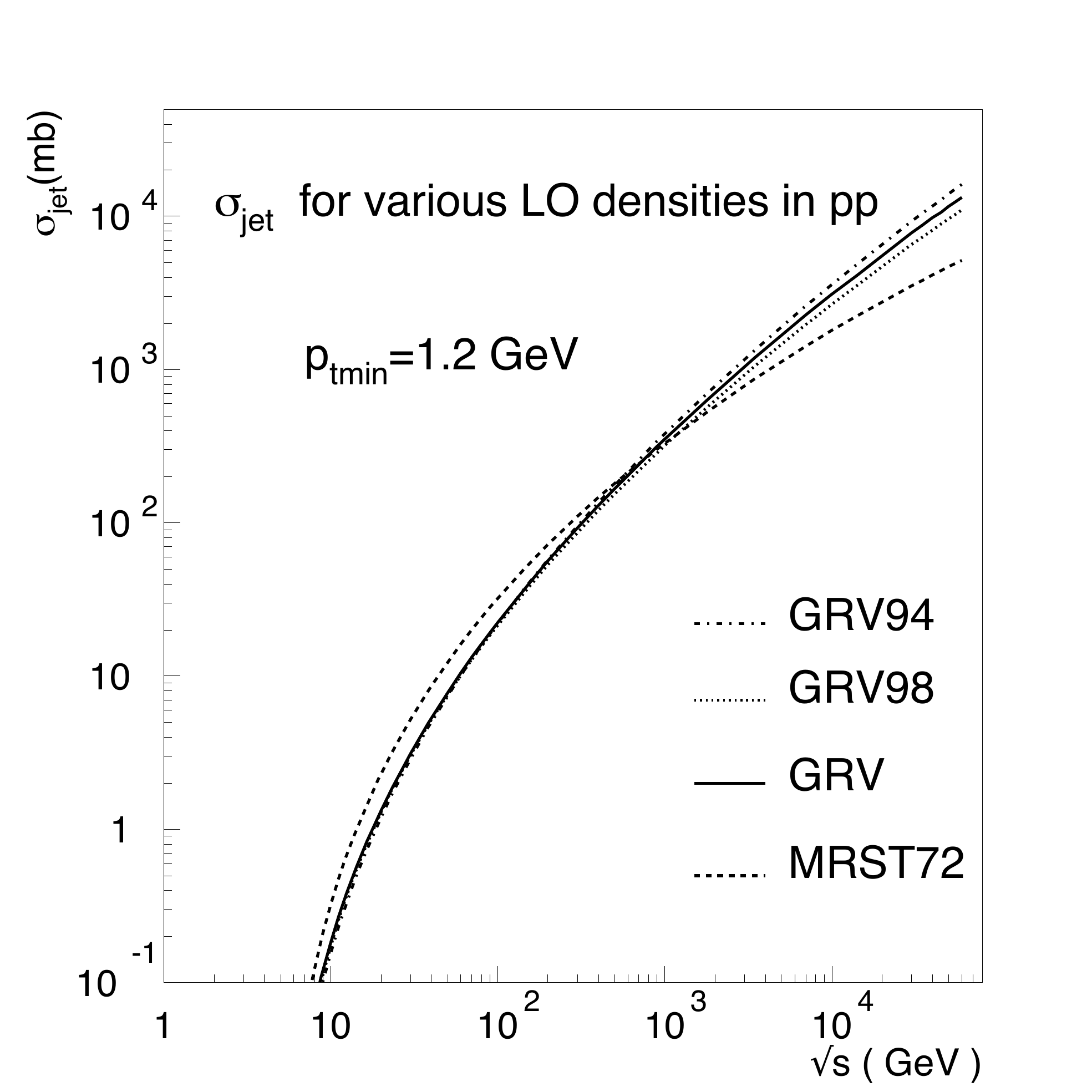}
\includegraphics{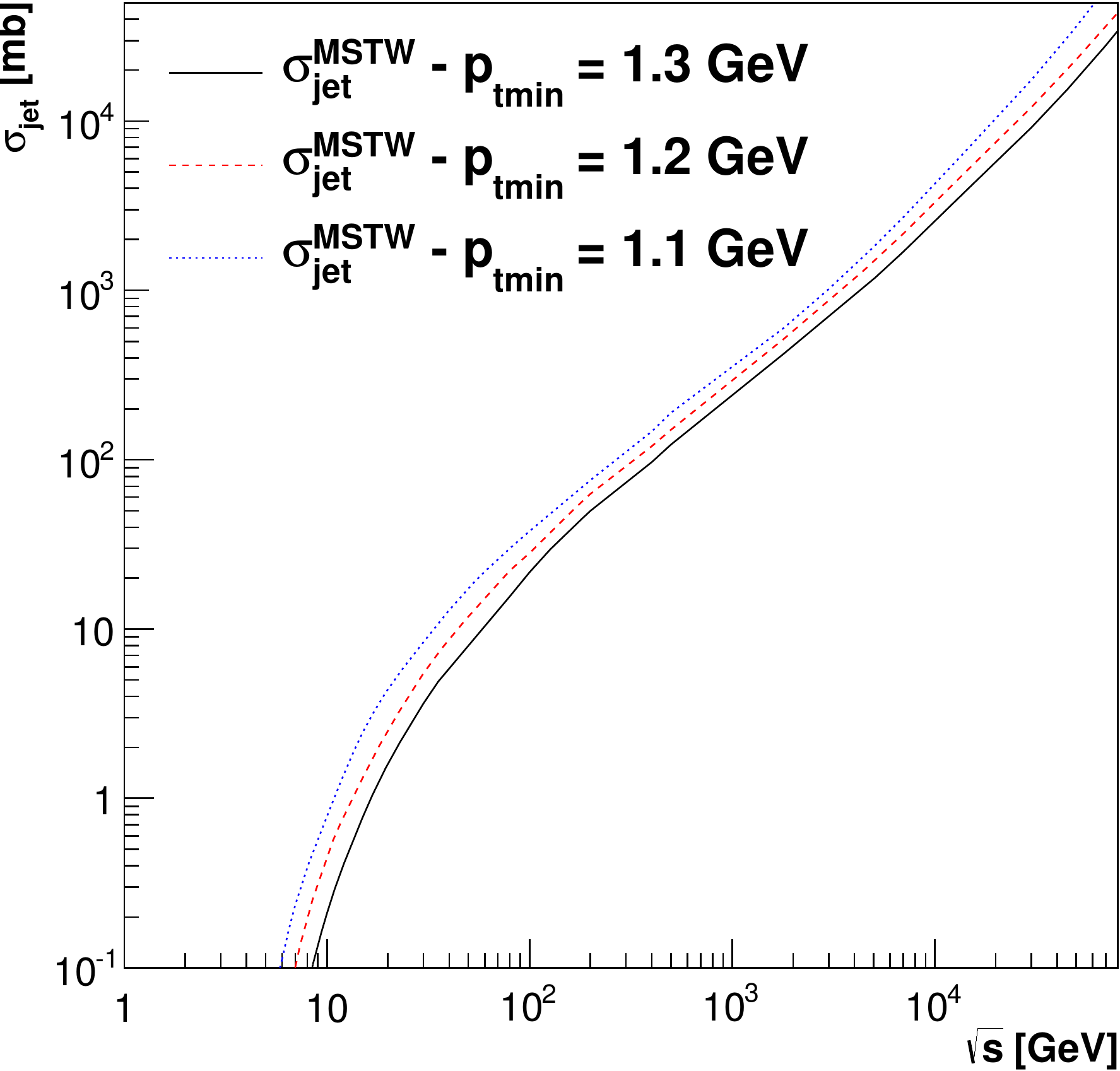}}
\caption{Integrated mini jet cross section for GRV (older) densities \cite{Gluck:1991ng,Gluck:1994uf,Gluck:1998xa} or  MRST72 \cite{Martin:1998sq}, and a fixed chosen 
$p_{tmin}$ value (left), and with MSTW2008 \cite{Martin:2009iq} LO PDFs (right) for  a range of $p_{tmin}$ 
values showing a distinct change of behavior, from a very fast rise to a power law, 
around $\sqrt{s}\simeq (10\div 20)\ GeV$. }
\label{fig:sigjet}
\end{figure}
Different densities  give rise to different energy dependence of the mini jet cross section, 
 as they differ in their very low-x behavior.
In Fig.~\ref{fig:sigjet}   we  display the energy -dependence of  $\sigma_{jet} (s)$, with parameter 
choices corresponding to $p_{tmin}\simeq 1\ GeV$ and different PDFs.
We use LO current PDFs
DGLAP evolved, from PDF libraries.  A clear 
threshold behavior appears around ISR energies and confirms the mini-jet explanation of the {\it soft edge}, 
that is, for $\sqrt{s} \gtrsim 10-20\  GeV$ the cross section rises mainly due to the energy evolution of the proton 
''radius",  coming from emergent gluon-gluon hard scatterings.

For a fixed $\ptmin$, these mini-jet cross sections rise 
as a power law, $s^{0.3-0.4}$,  and, 
when included in the formalism for the total cross section,  phenomenologically correspond 
to the hard Pomeron of Regge-Pomeron models.

For the transition to be at the $\sqrt{s}$ values as indicated by the    
rise of the cross section at ISR, the parameter $\ptmin$ has to be  roughly energy independent. 
This is in disagreement with other, typically Regge-Pomeron, formulations, but is  in agreement with the observation in  
\cite{Block:2014lna} and confirms the finding of our model, which, from the very beginning, used a 
constant $p_t$ value, labelled as $p_{tmin}$, as a divider between scattering giving rise to mini-jets 
and non-perturbative scattering.

However, this poses a serious problem with  the
rigorous limits imposed on cross sections by the Froissart bound. 
In various mini-jet or Regge-Pomeron based models,  this is   the main reason to introduce an 
energy dependent $\ptmin$. In so doing however, not only  there is no way to simply explain the 
{\it soft edge} as a fixed energy threshold, but  an important piece of physics remains hidden, 
namely soft gluon re-summation 
that tames the unphysical rise of the mini-jet cross section. To which we now turn.

\section{Soft gluon re-summation and the Froissart bound \label{Soft}}
We now turn to the {\it second fact} mentioned earlier. 
In our model the softening of the rise is obtained by soft gluon emission, 
a phenomenon 
always accompanying hard scattering, such as the one which results in the mini-jet production. 
Our starting point is that    resummed soft gluon emission  tames the rise of the mini-jet cross section.
Soft emission always decreases the observable cross section, since it introduces an overall transverse  momentum 
$K_t\neq 0$ in the center of mass of the colliding particles, namely it leads to acollinearity of the initial particles in the scattering. We have suggested that soft gluon QCD re-summation 
plays the same effect in hadron-hadron collisions, in particular accompanying  hard parton-parton scattering. 
In \cite{Godbole:2004kx}, we advanced the conjecture  that such phenomenon is behind the initial 
decrease of total cross sections, before the on-set of mini jets. Past ISR, perturbative QCD effects, 
i.e.  mini jets, take over, and the combination of the decrease due to soft gluon re-summation (SGR)  
and the rise of mini-jets  transforms the rapid power law rise into a 
milder behaviour, leading  to the second important  {\it fact} mentioned in Sect.\ref{soft-edge}. 

A quantitative description of this mechanism poses an extraordinary challenge, and we shall here 
briefly outline how we have approached the problem in our model \cite{Corsetti:1996wg}.
 
 We consider the following standard expression for soft $k_t$ re-summation, i.e.
\begin{eqnarray}
\frac
{d^2P({\bf K}_t)}
{d^2{\bf K}_t}=\frac{1}{(2\pi)^2}
\int d^2{\bf b} e^{i{\vecK}_t\cdot {\vecb}} e^{-h({\bf b})}\\
h({\bf b})= \int d^3 {\bar n}(k)[1-e^{-i \veck_t\cdot \vecb}]
\end{eqnarray} where  $d^3 {\bar n}(k)$ is the average number of soft quanta  
emitted during a  collision.
In  our model,   the Fourier transform of the above expression describes the impact parameter distribution 
of partons in the proton, $A(b,s)$, during the hard gluon-gluon collision. 
In the well known formulation of SGR \cite{Dokshitzer:1978yd,Parisi:1979se}, the lower limit of the $k_t$ integration 
is excluded, via an infrared cutoff of order $\Lambda_{QCD}$. This procedure is acceptable as long as there is 
no singularity in the infra-red region \cite{Parisi:1979se}, which is certainly not true when confinement effects play a role. 

Our model relies on  an adequate, albeit phenomenological,  inclusion of  {\it  the infra-red region}.  
We parametrize the very low $k_t$ emission with a parameter $p$, through an effect of zero momentum 
or close to zero momentum emission.  Our parametrization of the infra-red(IR) region has been discussed 
at length in many publications, see for instance  \cite{Grau:1999em} and  \cite{Godbole:2004kx}, and 
will not be repeated here, except to remind the reader that it basically amounts to describe the IR 
region through a singular but integrable expression for the coupling $\alpha_{eff}$ of very soft, infra-red gluons from  
the emitting quark current, namely we use
\begin{equation}
\alpha_{eff}=
\frac{12 \pi}{33-2N_f}\frac{p}{\log[1+p(k_t/\Lambda_{QCD})^{2p}]}
\end{equation}
The parameter $p$ regulates the singularity, it is by construction $1/2<p<1$. We determine its value 
phenomenologically, but we expect it to be related to the coefficient of the beta function.
With such an expression, $k_t$-resummation in transverse momentum  of soft gluons emission can be 
performed down into the  $k_t=0$ region. 

We label the impact parameter distribution accompanying mini-jet scattering 
with       the  subscript BN   as $A_{BN}^{pp}(p,PDF;b,s)$,    
pinpointing  to   the  need for  re-summation of  soft quanta  emitted in the so-called infrared catastrophe \cite{Bloch:1937pw}.  We have \bea
A_{BN}^{pp}(p,PDF;b,s)=\frac{
e^{-h(p;b,s)}}
{\int d^2 {\vecb} e^{-h(p;b,s)} }\\
h(b,s;p)=\frac{16}{3\pi} \int_0^{q_{max}}
\frac{dk_t}{k_t}
 \alpha_{eff}(k_t) \log\frac{2q_{max}}{k_t}[1-J_0(bk_t)] \label{eq:hdb} \\
\alpha_{eff}(k_t)\propto (\frac{k_t}{\Lambda})^{-2p} \ \ \ \ \ \ \ \ \ \  k_t\rightarrow 0
\eea
where the upper limit of integration $q_{max}$ indicates the PDF dependence, as we see shortly.
We have discussed the above  distribution in many publications, starting with  \cite{Corsetti:1996wg}. Its main characteristic is to 
include soft gluon re-summation down to $k_t=0$, and regulate the infrared singularity   through  a parameter $p$,
so as correspond to a dressed gluon potential $V¨\sim r^{2p-1}$ for $r\rightarrow \infty$. 
We have also shown an important consequence of  an expression such as the  
above for $\alpha_{eff}(k_t\rightarrow 0)$ \cite{Grau:2009qx}, namely  that, asymptotically,  
the regularized and integrated soft gluon spectrum  of Eq.~(\ref{eq:hdb}) is seen to rise as
\be
h(b,s;p)\rightarrow (b{\bar {\Lambda}})^{2p}
\ee 
 and the $b-$ distribution $A_{BN}^{pp}$ exhibits
a cut-off in $b-$space strongly dependent on the parameter $p$. 
When the above distribution is included in an eikonal mini jet model, such cut-off  is seen to give rise 
to a total cross section which rises asymptotically as $[\log s]^{1/p}$, as shown in  \cite{Grau:2009qx}, 
where details can be found.

\begin{figure}
\resizebox{1\textwidth}{!}{
\includegraphics{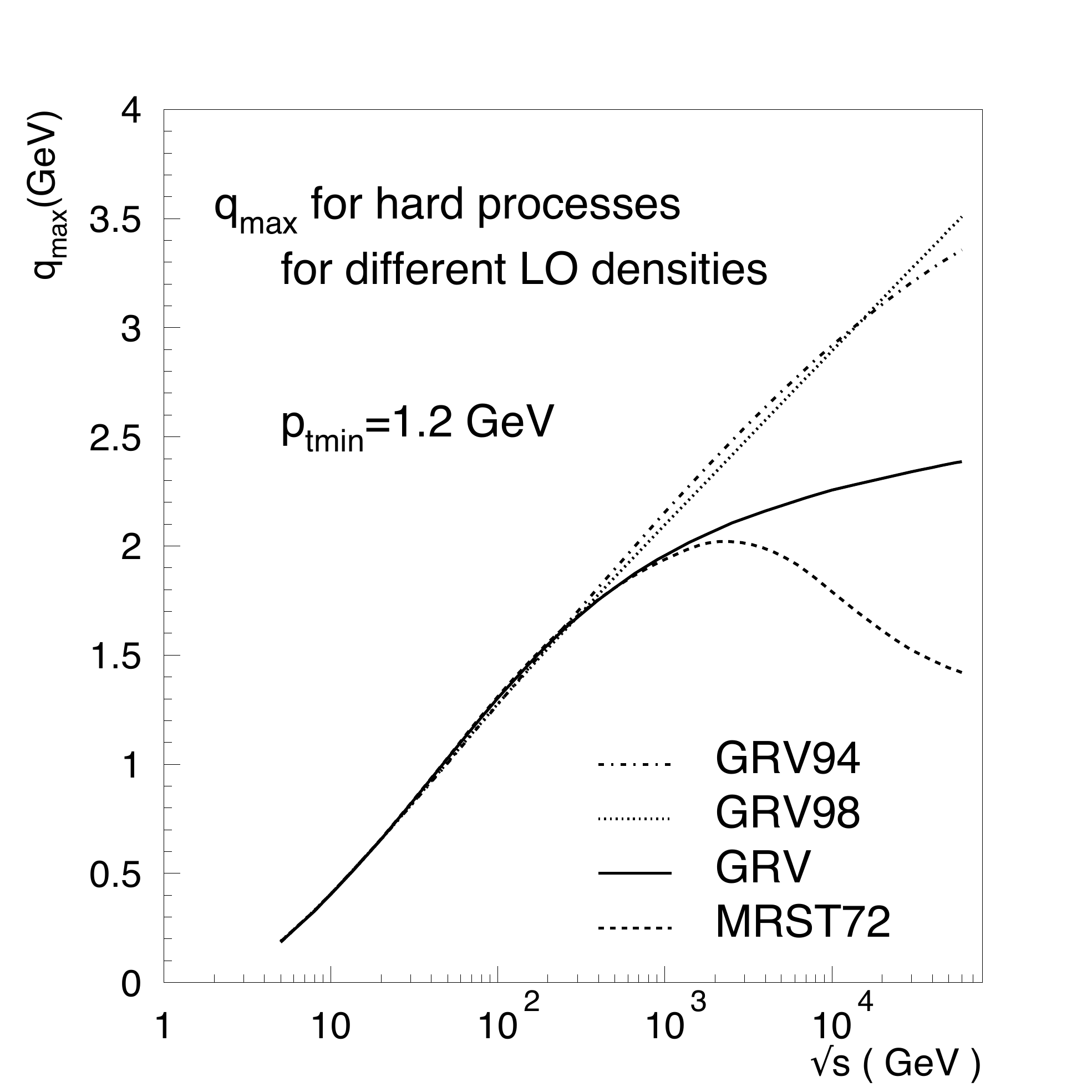}
\includegraphics{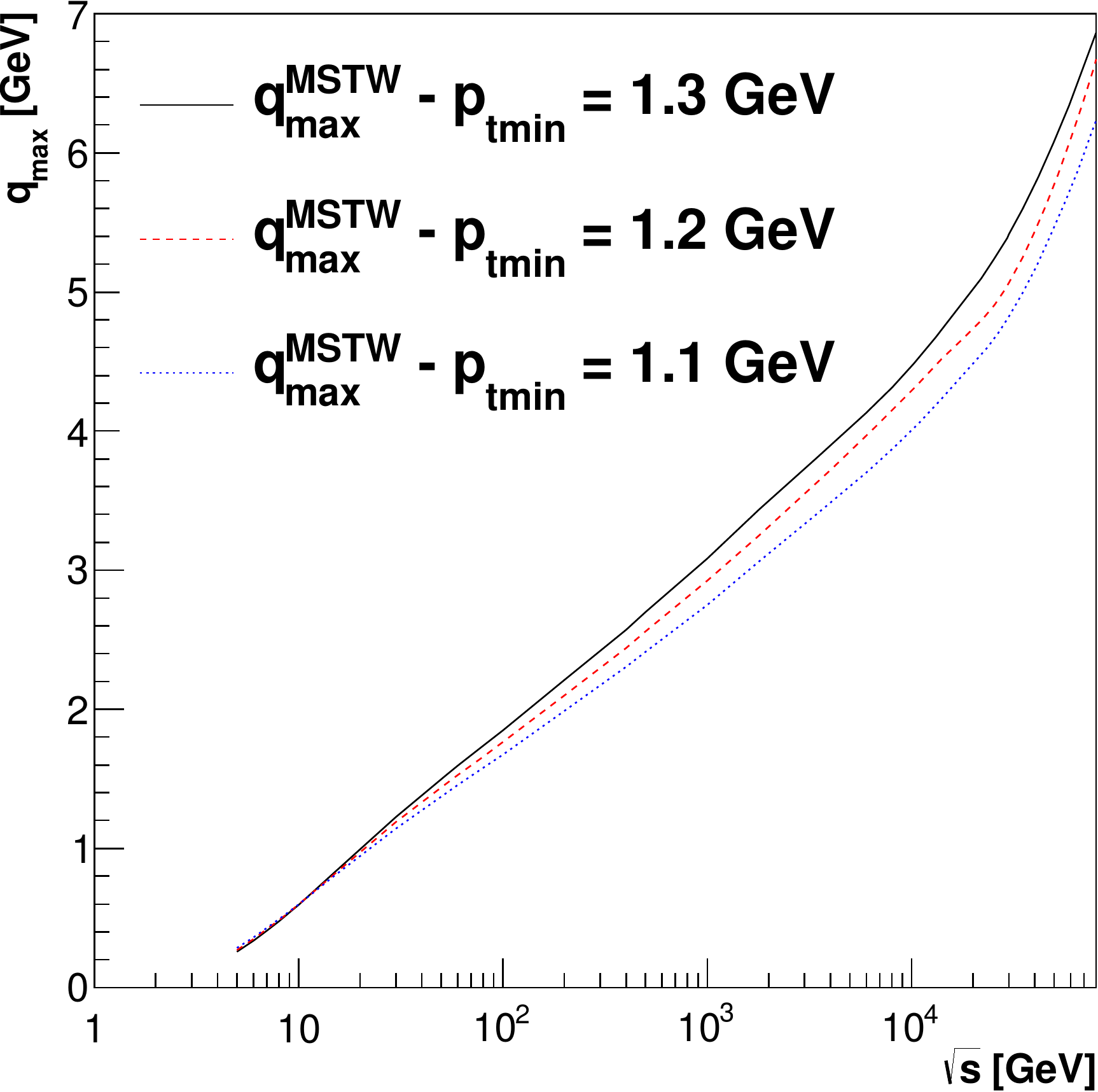}}
\caption{Dynamically generated scale $q_{max}$ for different PDFs  and a fixed  $p_{tmin}$ value (left panel) and MSTW2008 LO parton densities,  within a range of $p_{tmin}$ values.}
\label{fig:qmax}
\end{figure}

Phenomenologically, just as  the mini-jet cross section depends on the PDFs, so will   
the soft gluon spectrum accompanying the hard scattering and so will the  expression we propose for 
the impact parameter distribution of partons. This happens  through the maximum transverse 
momentum allowed to single  soft gluon emission during the hard gluon-gluon collision.  
The dynamical scale $q_{max}(s)$  in  Eq.~(\ref{eq:hdb})  is obtained from  the kinematics 
for single soft gluon emission \cite{Chiappetta:1981bw}.  In our model,   as discussed in the 
original formulation \cite{Grau:1999em}, we have made some approximations, such as that 
most contributions come from $p_t\simeq p_{tmin}$,  and have  then taken averages over the 
densities chosen for the mini-jet cross section.   Thus, we use   the 
{expression} 
\be
q_{\max } (s;p_{tmin})   = \frac{\sqrt {s}} {2}\, 
\frac{{\sum\limits_{i,j} {\int {\frac{{dx_1 }} {{x_1 }}\int
{\frac{{dx_2 }} {{x_2 }} \int_{z_{min}}^1 {dz f_i (x_1) f_j (x_2) \sqrt
{x_1 x_2 } (1 - z)} } } } }}
{{\sum\limits_{i,j} {\int {\frac{{dx_1 }} {{x_1 }}\int {\frac{{dx_2}}
{{x_2}} \int_{z_{min}}^1 {dz} f_i (x_1)f_j
(x_2) } } } }} \label{eq:qmax}
\ee
with $z_{min}=4p_{tmin}^2/(s x_1 x_2)$.\footnote{A typo in Eq.~\ref{eq:qmax} has been corrected wrt to the ArXiv v2 and published PRD version. }
This scale, the maximum transverse momentum  allowed for single soft gluon emission at a given energy,  
has a strong dependence on the PDFs used for its calculation, and on $p_{tmin}$, as well.  
In this paper we shall show results for  different LO PDFs. In previous publications \cite{Achilli:2011sw}, 
we   discussed the case of GRV \cite{Gluck:1991ng,Gluck:1994uf,Gluck:1998xa}, or MRST72 \cite{Martin:1998sq}.
 In the left panel of Fig.~\ref{fig:qmax} we show the $s$-dependence of the soft energy scale  $q_{max}$ 
 for these densities and for one fixed  value of $p_{tmin}$.
 New updated densities are now available, and we show in the right hand panel 
 the s- dependence 
 for a range of values for $p_{tmin}$ and for the recent LO PDF set, MSTW2008, for brevity   also called    MSTW  in this paper \cite{Martin:2009iq}. Thus,  while the scale 
 defining  the mini jet contribution, i.e. $p_{tmin}$, is fixed,   the model has a dynamically generated energy dependent scale, $q_{max}$.

\section{One-channel mini-jet model for  total, elastic and inelastic  cross sections}\label{sec:sigs}
The appearance of mini-jets in hadronic collisions has a long history \cite{Pancheri:1985ix}. The construction of the total hadronic cross section with mini-jets had 
started in  \cite{Gaisser:1984pg} with a simple model with an energy dependent $p_{T min}(s)$.    
 The original suggestion in \cite{Gaisser:1984pg} did not use the eikonal representation, and  was based on phenomenologically determined values of  $p_{T min}(s)$ at different energies. Embedding  the mini-jets in the eikonal   formulation \cite{Durand:1988ax}, unitarity would be automatically satisfied but the modeling  increased due to the request of specifying an  impact parameter distribution.  Durand and Pi's original suggestion and many others to follow, use  the one-channel description, but their use is limited and in what follows  we shall clarify these limitations. 
 
 Presently,  good descriptions of all the components of the cross section are obtained through multi-channel formulations.
 Some of them \cite{Khoze:2013jsa} are discrete {\it \'a la Good and Walker} \cite{Good:1960ba}, others are based on a  continuous \cite{Lipari:2009rm,Lipari:2013kta}.  The  price to pay then  is  addition of  more free parameters. 
 Current models in perturbative QCD,   on accounting for multiple Pomeron interactions,  invoke enhanced ``fan'' diagrams \cite{Ryskin:2012az,Gotsman:2012rq,Ostapchenko:2010vb} which require the knowledge of non-linear coupling between Pomerons. Non-linear equations such as the Balitsky-Kovchegov  (BK) equation \cite{Balitsky:1995ub,Kovchegov:2001ni},
are used to  describe these non-linear contributions to diffraction. Such non-linear couplings become important when the gluon momentum becomes very small, a region insofar unreachable by perturbative QCD. Thus,  while diffraction needs a multichannel approach, its complete description and phenomenology so far have to rely on more  parameters \cite{Ostapchenko:2010gt}, some of which embody the non-linear behavior in the unknown infrared region. 

In this  section, we shall discuss the mini-jet  
contribution to total, elastic and inelastic cross section, using a  one-channel eikonal formulation, and leave to a 
forthcoming publication, the implementation to a general  model and the inclusion 
of diffractive processes. 

\subsection{The total cross section}
  In this section, we shall    update     our mini jet model results to LHC13 energies and beyond. 
  To construct the total cross section, mini-jets are embedded into the eikonal formulation. Starting with
 \be
\sigma_{total}=2 \int d^2 {\vec b} [1-\Re e( e^{i\chi(b,s)})]\label{eq:sigtot}
\ee
and neglecting the real part in the eikonal at very high energy, the above expression further simplifies into
\be \sigma_{total} =2 \int d^2 {\vec b} [1-e^{-\chi_I(b,s)}] \label{eq:onechtot}
\ee
where $\chi_I(b,s)=\Im m\chi(b,s)$.  Notice that $\Re e\chi(b,s)\simeq 0$ is a reasonable  approximation 
for the scattering amplitude in $\vecb$-space at $t=0$, where very large values of the impact parameter dominate 
and  phenomenologically     the ratio of the real to the imaginary part of the forward scattering amplitude $\rho(s) << 1$. 
By properly choosing a function $\chi_I(b,s)$, all  total hadronic cross sections, $pp$, $p{\bar p}$, $\pi p$, etc.,   can  be described  up to 
currently available data \cite{Grau:2010ju}. In the vast majority of  models, new data have often  required  an adjustment of the 
parameters which give $\chi_I(b,s)$. 

In previous publications,
we had proposed a band whose upper border gave a good 
prediction for LHC results.  By updating the model and anchoring the parameter set to 
LHC results, one can now proceed to refine our predictions for higher energies, LHC13 and beyond to the cosmic rays region.
  The eikonal function of the mini-jet model of \cite{Godbole:2004kx,Grau:1999em} 
is given by
\be
2\chi_I(b,s)=n_{soft}^{pp}(b,s)+n_{jet}^{pp}(b,s)=A_{FF}(b)\sigma_{soft}^{pp}(s)+A_{BN}^{pp}(p,PDF;b,s)\sigma_{jet}(PDF,p_{tmin};s)\label{eq:chi}
\ee
The first term  includes collision with $p_t \le p_{tmin}\sim (1\div 1.5 )\ GeV$, the second is
obtained from the mini-jet cross section. The term $n_{soft}^{pp}(b,s)$ is not predicted by our model so far 
and we parametrize it here with $\sigma_{soft}^{pp}(s)$,  obtained  with a constant and one or more decreasing 
terms, and $A_{FF}$, the impact parameter distribution  in the non perturbative term, obtained through a  
convolution of two proton form factors.

As we have seen,  the second term  in Eq.(\ref{eq:chi}) 
   is numerically
 negligible at energies $\sqrt{s}\lesssim 10 \ GeV$. The perturbative, mini-jet, part discussed previously  
 is  defined  with $p_t^{parton}\ge p_{tmin}$ and is determined through a set of perturbative parameters 
 for the jet cross section, namely a choice of  PDFs  and the appropriate  $p_{tmin}$. Since  soft gluon 
 re-summation includes all order terms in soft gluon emission,  our model   uses  LO,   library distributed,  PDFs. 

\begin{figure}[h]
\centering
\resizebox{0.8\textwidth}{!}{
\includegraphics{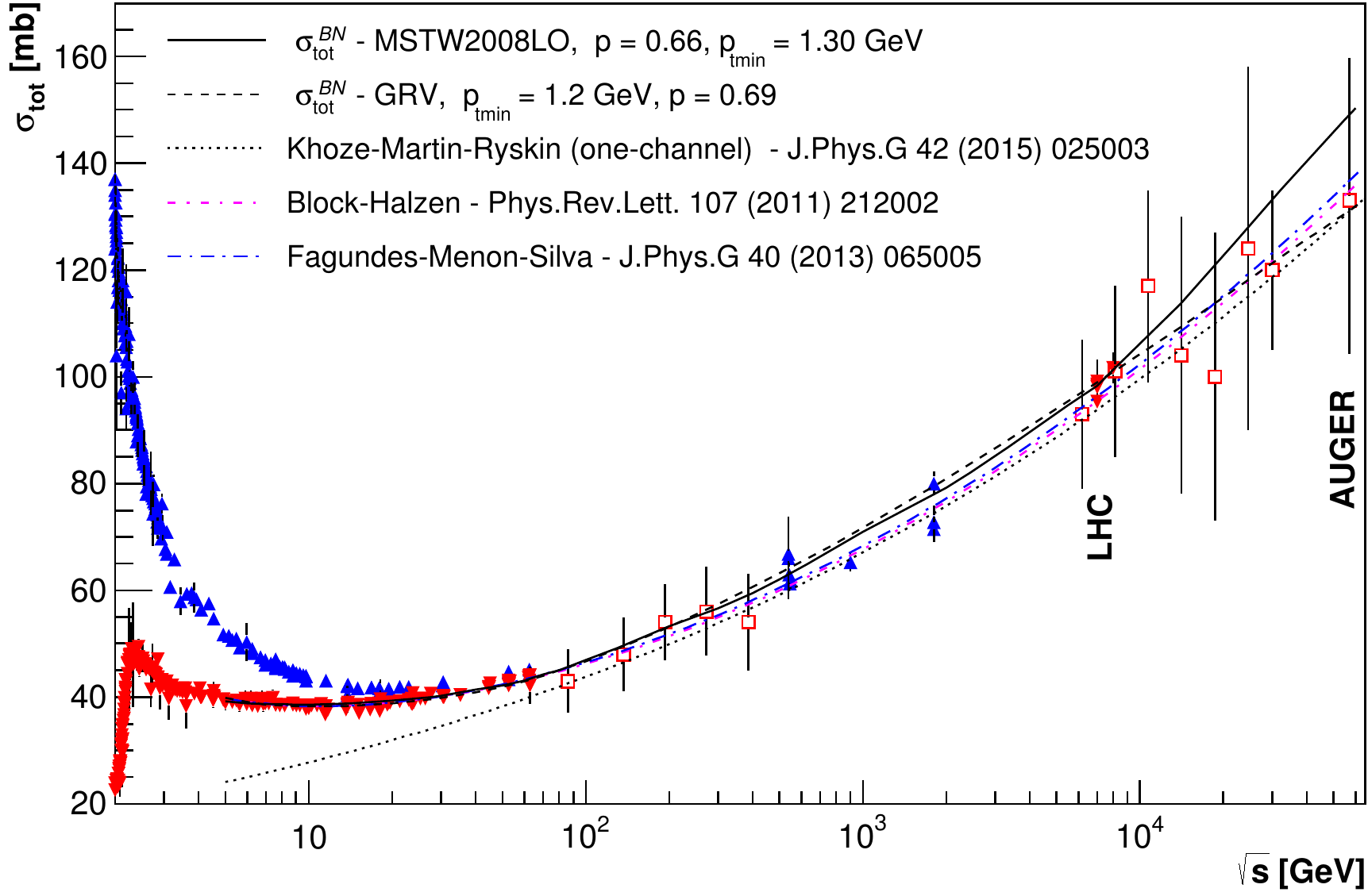}}
\caption{ QCD mini-jet with soft gluon resummation model and $pp$ total cross section.
Accelerator data at LHC include TOTEM \cite{Antchev:2013paa,Antchev:2013iaa} and 
ATLAS measurements \cite{Aad:2014dca}. 
  Neither  ${\bar p} p $ nor the cosmic ray data points 
 have   been used for the calculation.
 Our results are shown with predictions in the high cosmic ray region obtained with two   types of densities, 
 GRV and MSTW, and they are compared with  one- channel model 
  from Khoze et al.\cite{Khoze:2014nia}. The  red {\it dot-dashed} curve corresponds to fits to the total cross section by Block and Halzen \cite{Block:2011vz}, the dot-dashed blue line represents the fit by Fagundes-Menon-Silva \cite{Fagundes:2012rr}.}
\label{fig:pptot}
\end{figure}

In our previous publications \cite{Achilli:2007pn,Achilli:2011sw}, we have reproduced  data for $pp$ and ${\bar p} p$, 
up to the TeVatron results.   
However,  the large differences among the  Tevatron measurements  
did not allow a precise description at higher energies, such as those at  LHC. Therefore we have updated 
our analysis,   using only $pp$ data,  ISR and the recent LHC measurements, and including  a more    recent  set of LO densities, 
MSTW2008 \cite{Martin:2009iq}.
The  values of $p$ and $p_{tmin} $ which better reproduce the LHC result are obtained by varying 
$p_{tmin} \simeq 1\div 1.5\ GeV$ and $1/2\lesssim p \lesssim0.8$. The result,  for  the total $pp$ cross sections, 
is shown in Fig.~\ref{fig:pptot}.  The    ${\bar p} p $ points are shown, but  have  not been used for the phenomenological fit. 
Cosmic ray extracted values for $pp$ have not been used either.
We  have included the curves corresponding to GRV densities and compared our model  with  other predictions  \cite{Block:2011vz,Khoze:2014nia,Fagundes:2012rr}. Table \ref{tab:total} contains the points corresponding to our model results for both GRV and MSTW2008 densities.\footnote
{In the arXiv v2 and in the published PRD paper, there were errors in the table for MSTW predictions at 13 and 57 TeV. These have now been corrected.}
Results for MRST72 densities can be found in \cite{Fagundes:2014fza}, together with  details of  different parameter sets used for the different PDFs.
\begin{table}[H]
\caption{Total cross section values in mb, from the mini-jet model with two different PDFs sets. }\label{tab:total}
\centering 
\begin{tabular}{|c|c|c|}
\hline
$\sqrt{s}$ GeV&$\sigma_{total}^{GRV}$ mb &$\sigma_{total}^{MSTW}$ mb\\ \hline
5		&	39.9	   &39.2	          	 \\
10		&	38.2	   &	38.6          \\
50		&	41.9	   &	 42.2         	 \\
500		&	63.2	   &	62.0	          \\
1800	&	79.5	   &	 77.5        \\
2760      &     85.4   &    83.6         \\
7000	&	98.9	   &	  98.3       \\
8000	&	100.9 & 101.3		        \\
13000	& 108.3	& {111.7}\\
14000	&	109.3  &113.7         \\
57000	&	131.1	&{149.2}\\
\hline
\end{tabular}
\end{table}
We notice that our model is able to describe very well all the total cross section  accelerator data,  and gives  a good agreement with cosmic ray  data.  The AUGER point falls within the two different paramerizations we are using, full line for MSTW and dashes for GRV. By construction, both parametrizations remain very close up to  LHC7 and LHC8  energies, and start diverging as the energy increases, as a consequence of  the uncertainty on the very low-x behavior of the densities.

To summarize the results of this section, in the model we have proposed, past ISR energies mini-jets appear as  hard gluon-gluon collisions accompanied by soft gluon emission $k_t$-resummed down into the infrared region. In this language, we have a {\it dressed} hard scattering process, with the mini-jet cross section giving the same energy behavior as the hard Pomeron,  and soft gluon resummation providing {\it the dressing},  in which the hard interaction is embedded. The eikonal formulation then transforms this {\it dressed hard gluon } interaction into a unitary ladder.  The main difference with other mini-jet models such for instance in  \cite{Giannini:2013jla}, is the taming mechanism  ascribed to soft gluon resummation in the infrared region.

\subsection{One-channel eikonal mini jet models and  the  inelastic cross section}
The inelastic total cross section is defined by subtraction  from  the total and the elastic cross sections.   However, experimentally, it is usually defined only in specific phase space regions, and eventually extrapolated via MC simulation programs, which also include parameters and choice of models in the diffractive region. {One exception is TOTEM which covers a large rapidity range.} In this subsection, we shall focus on one,  theoretically well defined, part of the inelastic cross section, what we define as {\it uncorrelated}, which is appropriately described in the mini-jet context and through the one-channel mode. In the following we shall see how. 

Since our study \cite{Achilli:2011sw} on the inelastic cross section at LHC, soon followed by the first 
experimental results \cite{Aad:2011eu}, data related to measurements in different kinematic regions have appeared. 
Extensive and detailed measurements have been obtained  for  the inelastic proton-proton cross section by  CMS \cite{Chatrchyan:2012nj}, ATLAS \cite{Aad:2011eu}, TOTEM \cite{Antchev:2013haa,Antchev:2013iaa}, ALICE \cite{Abelev:2012sea}  and LHCb \cite{Aaij:2014vfa} Collaborations. These measurements cover different regions, central and mid-rapidity, large rapidity, high and low mass diffractive states. 
Extensive QCD modeling, including minijets \cite{Ostapchenko:2014mna,Ostapchenko:2005nj,Kohara:2014cra,Goulianos:2014hqa}, goes in describing the different regions.

Here, we 
 consider the implication of any given one-channel eikonal model. 
Thus, we  repeat the argument about the relation  between the Poisson 
distribution of independent collisions and diffractive processes given in \cite{Achilli:2011sw},
where we stressed that the inelastic cross section 
in a one-channel eikonal model coincides with the 
sum of independently  (Poisson) distributed collisions  in b-space. Namely, with
\be
\sigma_{total}=\sigma_{elastic}+\sigma_{inel}\label{eq:sigonech}
\ee
then, in a one-channel ({\it one-ch})  mode,  
\be
\sigma_{inel}^{one-ch}\equiv \sigma_{tot}-\sigma_{elastic}^{one-ch}=\int d^2 {\vecb} [1-e^{-2 \chi_I(b,s)}]\label{eq:siginelonech}
\ee
But since 
\be
\sum_1^{\infty} \frac{({\bar n})^n e^{-{\bar n}(b,s)}}{n!}=1-e^{-{\bar n}(b,s)}\label{eq:poisson}
\ee
one can  identify the integrand at the right hand side of Eq.~(\ref{eq:siginelonech}) with a sum of  totally independent collisions, with $2 \chi_I(b,s)={\bar n}(b,s)$.  
We suggest that this means that in so doing one  excludes 
 diffraction and other quasi-elastic processes 
 from the integration in Eq.~(\ref{eq:siginelonech}). Hence, the  simple splitting of the total cross section as in  Eq.~(\ref{eq:sigonech}) needs to be better qualified 
when a one-channel  eikonal is used. In such a case, the ``elastic'' cross section 
\be
\sigma_{elastic}^{one-ch}=
\int d^2 {\vec b} |1-e^{-\chi_I(b,s)}|^2 \label{eq:sigel-onech}
\ee
must be including  part of the inelastic 
contribution, i.e. 
\be
\sigma_{elastic}^{one-ch}=\sigma_{elastic}+ \ diffractive\  or \ otherwise \ correlated \ processes
\ee
and $\sigma_{inel}^{one-ch}$  is only the non-diffractive part.  Within this approach, we can compare Eq.~(\ref{eq:siginelonech}) with data.

\begin{figure}[H]
\centering
\resizebox{0.8\textwidth}{!}{
\includegraphics{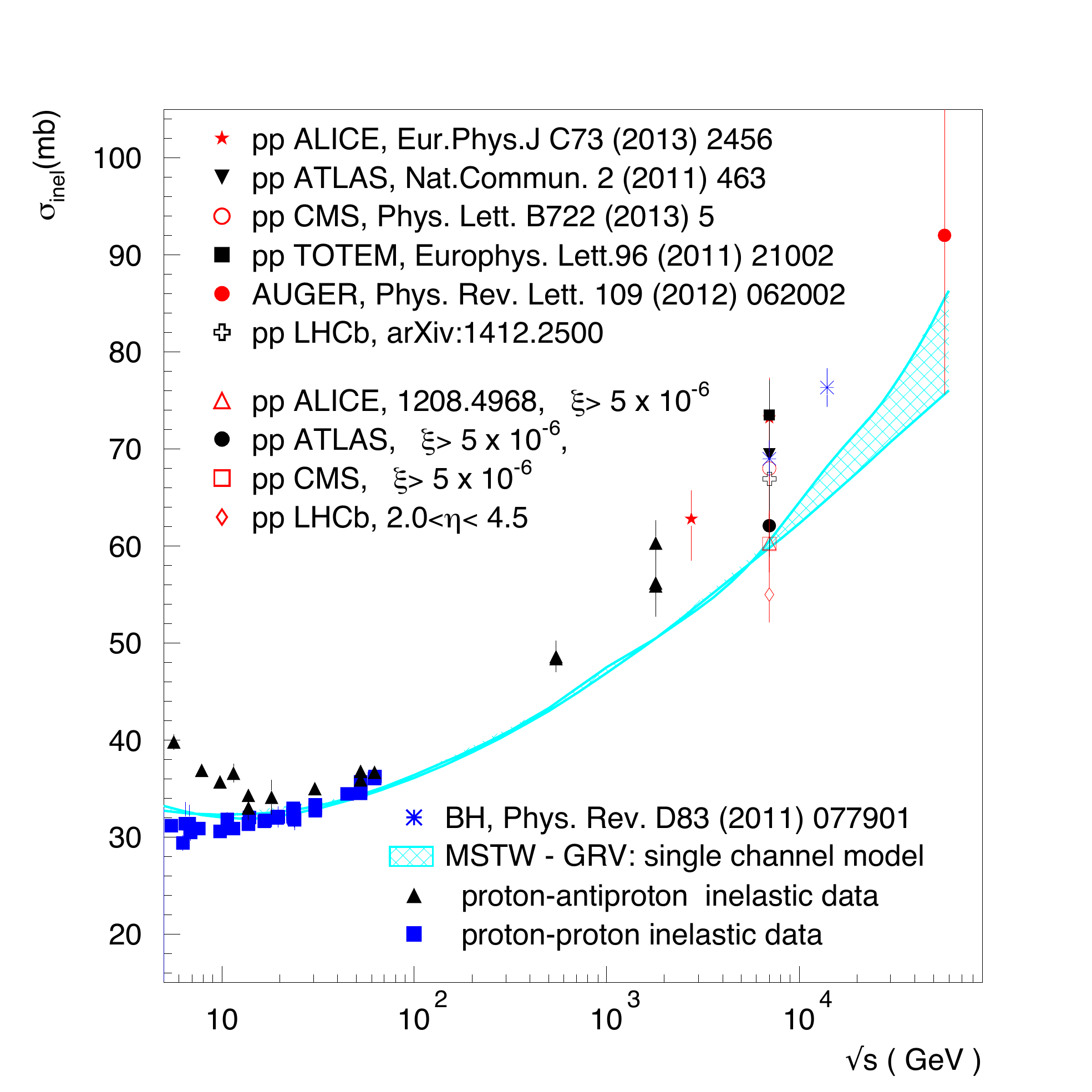}}
\caption{Data for the inelastic cross section and comparison with  GRV and MSTW2008 densities in the present  one-channel mini-jet model. We also show comparison with  Bloch and Halzen (BH) results  \cite{Block:2011uy}.}
\label{fig:olga-inel}
\end{figure}

This comparison is shown in  Fig.~\ref{fig:olga-inel}, where   the present  inelastic cross section data up to AUGER energies \cite{Collaboration:2012wt} are plotted. The blue band corresponds to the expectations from     Eq.~(\ref{eq:siginelonech}) where the same eikonal function $\chi_I(b,s)$  which gives the total cross section of Fig.~\ref{fig:pptot} is used. Having anchored the eikonal $\chi_I(b,s)$ to the LHC total cross section,  the band indicates  the  spread of predictions  due to the different asymptotic low-x behavior of the employed densities,  as the energy increases beyond LHC8. The top curve corresponds to MSTW2008, the lower one to GRV.

The comparison with experimental data is very interesting. While
   the present LHC inelastic cross section data  span a range of values corresponding to different kinematic regions,  Eq.~(\ref{eq:siginelonech}) identifies  the region where uncorrelated events described by mini-jet collisions, parton-parton collision with $p_t>p_{tmin}$, play the main role. From the comparison with data, we can identify it with  the region  $\xi=  M^2_X/s \ge 5\times  10^{-6}$ where  three LHC experiments, ATLAS \cite{Aad:2011eu}, CMS \cite{Chatrchyan:2012nj}  
and ALICE \cite{Abelev:2012sea}, agree to a 
common value within  a small  error.
 This measurement is in the  high mass region (for instance, at LHC7 the lower 
bound gives $M_{X}=15.7 \ GeV$).  LhCb results correspond to a lower cross section, but they  do not cover the same region of phase space.

\begin{figure}[H]
\centering
\resizebox{0.8\textwidth}{!}{
\includegraphics{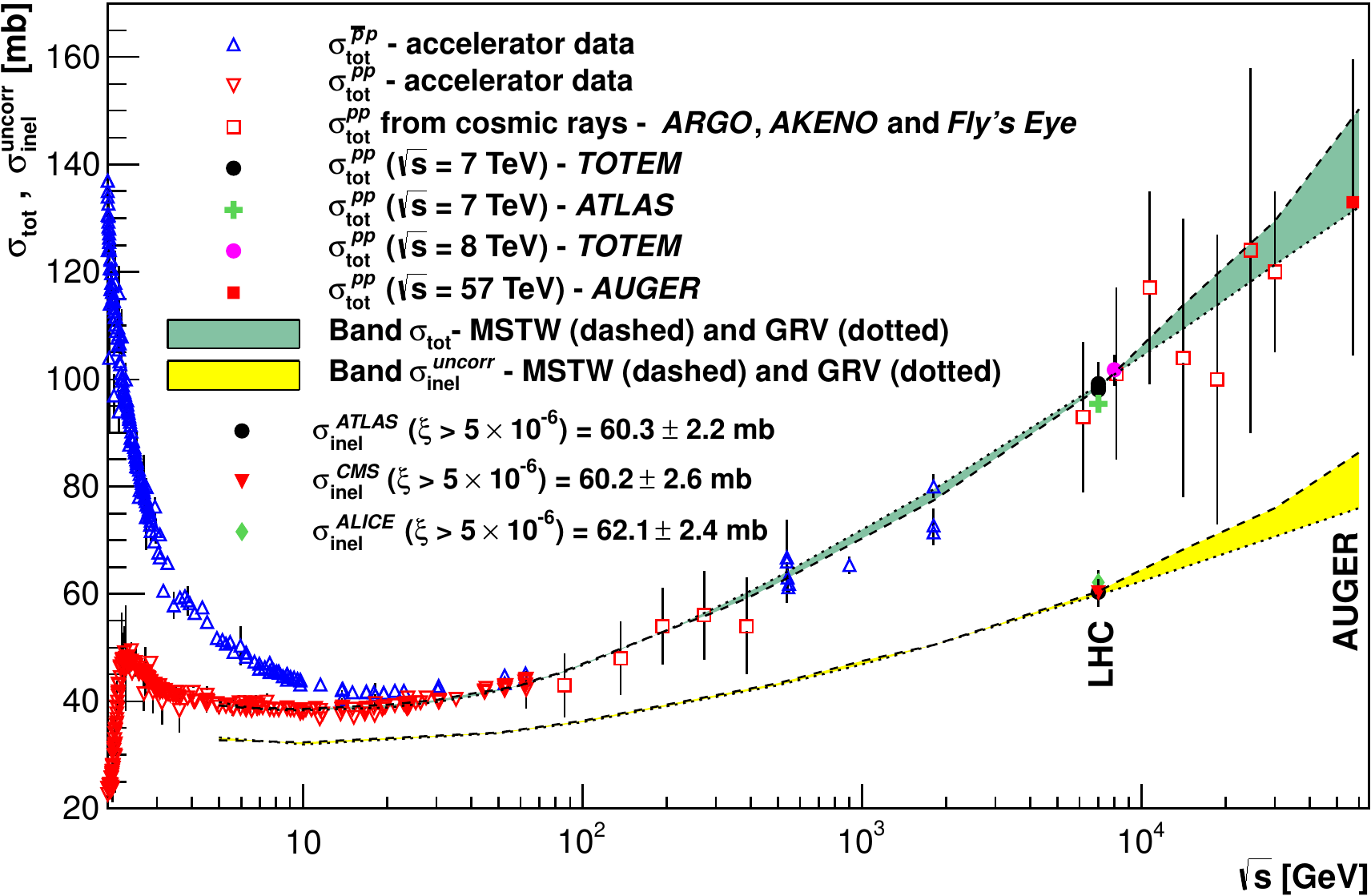}}
\caption{Results for our eikonalized QCD mini-jet with soft gluon resummation model are shown for  $pp$ total cross section and for the inelastic uncorrelated part of the inelastic cross section.  
Accelerator data at LHC include TOTEM \cite{Antchev:2013paa,Antchev:2013iaa} and 
ATLAS measurements \cite{Aad:2014dca}.
The inelastic uncorrelated cross section is compared with inelastic processes 
for $M^2_X/s > 5 \times 10^{-6}$ as measured  by ATLAS \cite{Aad:2011eu}, CMS \cite{Chatrchyan:2012nj}  and 
ALICE \cite{Abelev:2012sea}.  The green and yellow bands give the uncertainty following  
use of different PDFs sets, MSTW2008 and GRV,
as also shown in   \cite{Fagundes:2014fza}.}
\label{fig:ppall}
\end{figure}

The results of this and of the previous subsection are summarized in  
 Fig.~\ref{fig:ppall} where the bands  correspond to different PDFs used in the calculation of mini-jets and to their different extrapolation to very low-x at the cosmic ray energies.

The dashed yellow band is the one-channel inelastic cross section that only includes Poisson-distributed
independent scatterings. That is, once the parameters of the eikonal $\chi_I(b,s)$ are chosen to give an optimal reproduction of the  the total cross section, the computed
inelastic cross section immediately gives the uncorrelated part of the total inelastic cross section.
The importance of this fact for cosmic ray deduced
 $pp$ cross sections has been noticed in \cite{Fagundes:2014fza} and shall be examined further elsewhere.

\subsection{Diffractive, elastic and inelastic cross sections reexamined}
The total cross section, which our model successfully describes, includes different components, but only  one 
of them 
is well defined  experimentally as well as theoretically,  that is the elastic cross section.
It is  well known that  one-channel eikonal models fail to simultaneously describe the total and the elastic cross section 
 through the entire available CM energy range, with the same parameter set. In the last sub-section, we have delineated this shortcoming through the observation 
 that once minijets become operative  past the {\it soft edge},  the computed elastic cross section includes correlated inelastic collisions  and the computed inelastic 
 lacks the same (i.e., its correlated inelastic part). We now discuss this matter in detail so as to make these 
 statements quantitative. We shall do so through  the one-  channel mini-jet model with a suitable parametrization of diffractive data.

In one- channel eikonal models, with the inelastic part given by  Eq.~(\ref{eq:siginelonech}), the elastic  part of the total cross section is given by Eq.~(\ref{eq:sigel-onech}). Notice that  
whereas Eq.~(\ref{eq:siginelonech}) is exact, in Eq.~(\ref{eq:sigel-onech}) the real part of the eikonal function 
has been neglected, as in Eq.~(\ref{eq:sigtot}). 

\begin{figure}[h]
\resizebox{1.1 \textwidth}{!}{
\includegraphics{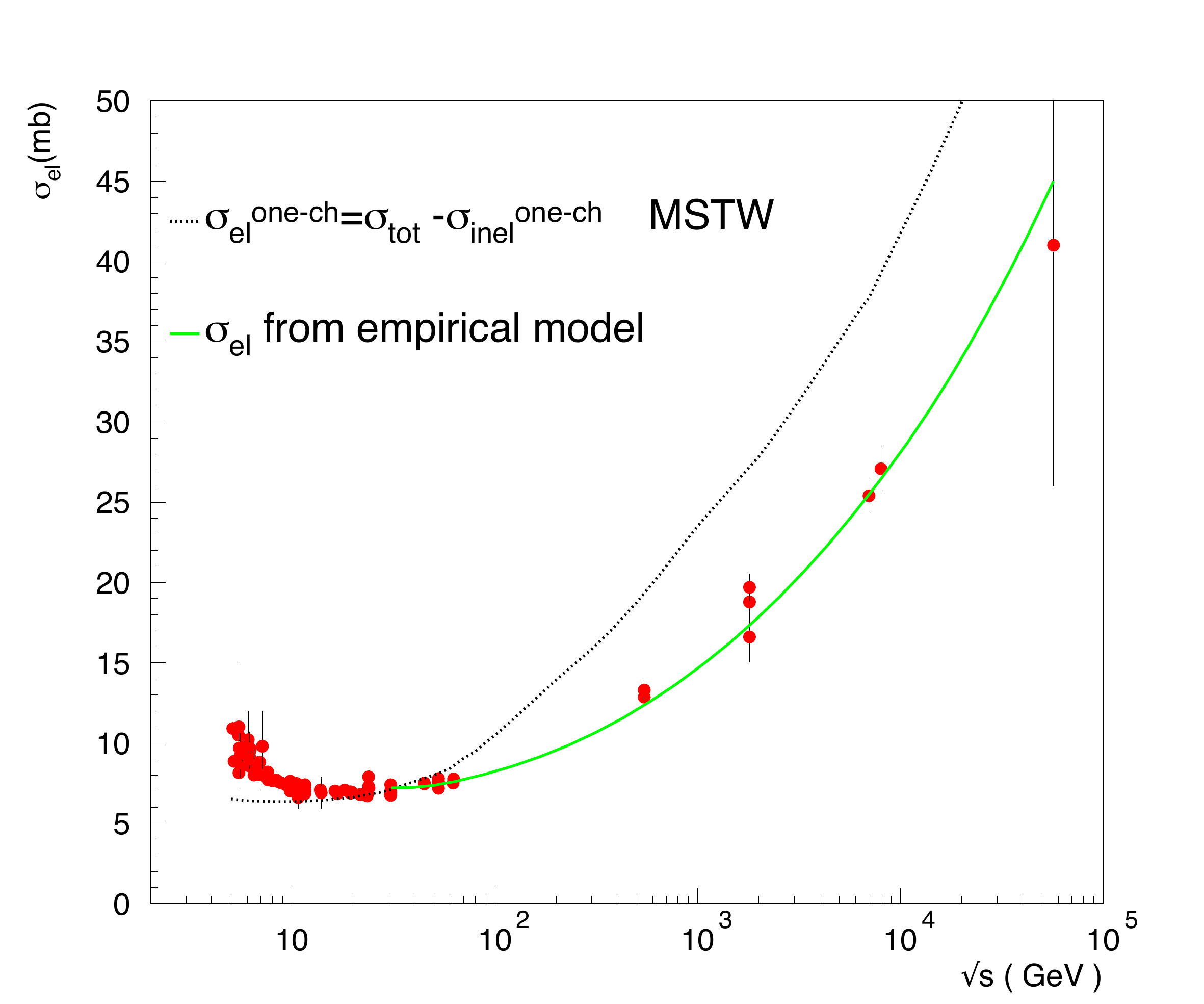}
\includegraphics{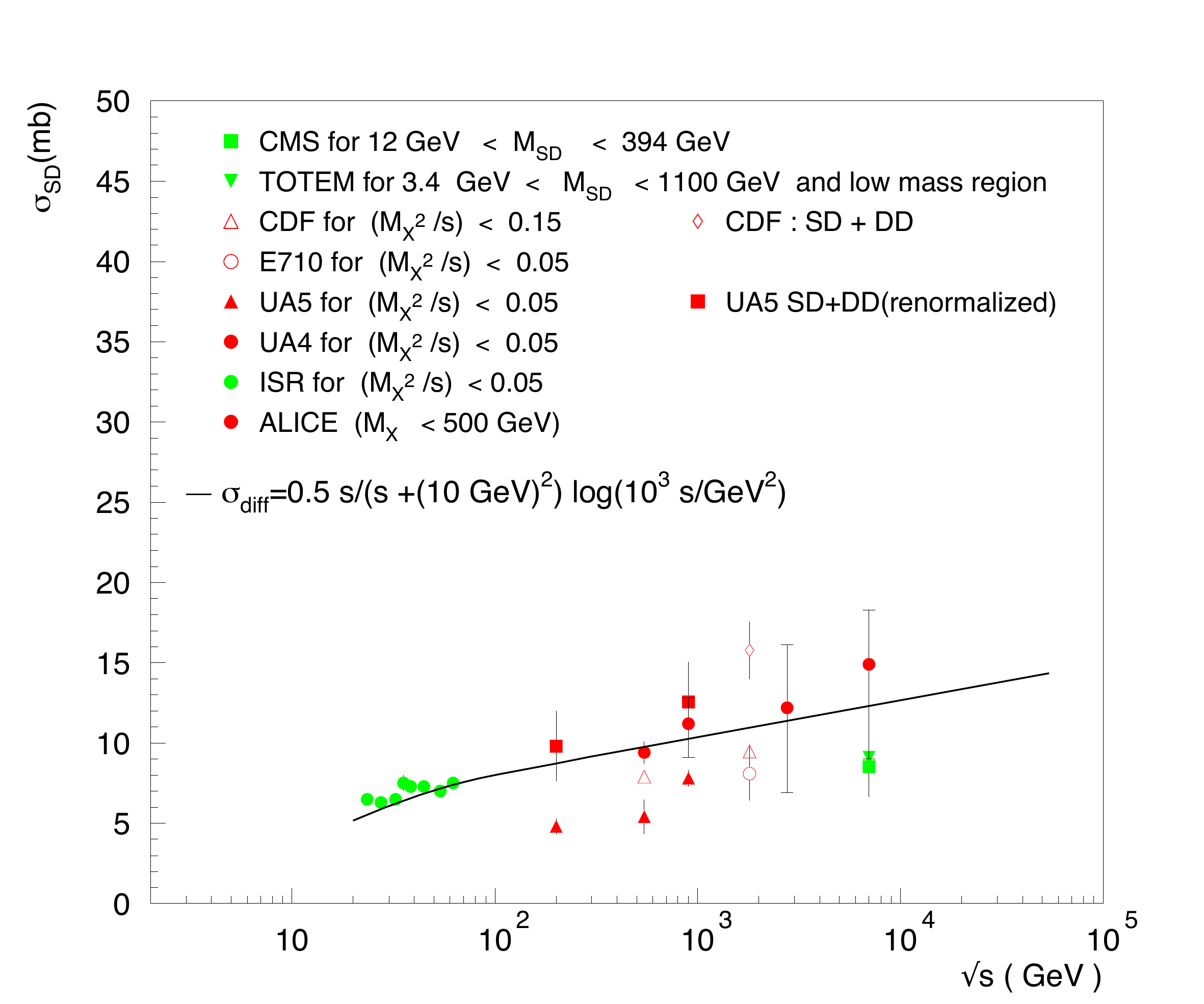}}
\caption{At left, we show the elastic $pp$ cross section from  the one-channel mode given by the dotted curve, with  choice of MSTW2008 PDF as in the upper curve of Fig.~\ref{fig:pptot}. The    green curve corresponds to an empirical parametrization of all differential elastic $pp$ data \cite{Fagundes:2013aja}. Comparison is done with both $pp$ and $p{\bar p}$ data. The right hand panel shows  diffraction data from E710 \cite{Amos:1992jw}, UA5 \cite{Ansorge:1986xq,Alner:1987wb},  UA4 \cite{Bernard:1986yh}, ISR \cite{Armitage:1981zp}, CDF \cite{Affolder:2001vx}, CMS \cite{Dutta:2014wda}, TOTEM \cite{Antchev:2013iaa}  and ALICE \cite{Abelev:2012sea}  compared with  the  parametrization given by  Eq.~(\ref{eq:sigdiff-EU})  mentioned in the text. }
\label{fig:elastic-onech-diff}
\end{figure}

Eq.~(19) reproduces with a good approximation the elastic cross section data up to the onset of minijets, deviating significantly from the data already at energies ~100 GeV. In particular, at the  Tevatron, 
Eq.~(\ref{eq:sigel-onech}) gives an elastic cross section roughly 30 \% higher than the data. This 
is shown in  the left hand plot of Fig.~\ref{fig:elastic-onech-diff}, where the one-channel result from Eq.~(\ref{eq:sigel-onech})  is plotted together with elastic scattering data and an empirical parametrization of all elastic differential cross section  $pp$ data from ISR to LHC7 \cite{Fagundes:2013aja}. This parametrization leads to  the expression
\be
\sigma_{elastic} = At_{0}e^{Bt_{0}}E_{8}(Bt_{0})+\frac{C}{D}+2\sqrt{AC}\cos\phi
t_{0}e^{(B+D)t_{0}/2}E_{4}\left(\frac{(B+D)t_{0}}{2} \right), \label{eq:sigelemp}
\ee
where
\be E_{n}(x) = \int_{1}^{\infty}\frac{e^{-xy}}{y^{n}} dy.
\ee
The model of \cite{Fagundes:2013aja} is based on the well known  Phillips and Barger model \cite{Phillips:1974vt} for the elastic differential cross section, implemented by a form factor term to fully reproduce the optical point, and hence the total cross section, as well as the forward slope. Through suitable predictions for the high energy behavior of the parameters,  s
the parameterization of \cite{Fagundes:2013aja}  provides  a model independent prediction  both for  elastic and total cross sections at very high energies, and hence can  be used as a good test of different models 
in the high energy region beyond present accelerator data.

The left hand plot of Fig.~\ref{fig:elastic-onech-diff}  shows that  at low energies, before the onset of mini-jets, one- channel models may be used to describe both elastic and total cross sections. However,  past  ISR energies 
the  threshold  of perturbative QCD, reflected in the appearance of the  {\it soft edge}, is crossed, 
and  one-channel models fail.  One-channel models are also unable to reproduce the behaviour of the 
differential elastic cross section, and multichannel models with added parameters are then 
needed to describe diffraction. The difficulty with proper descriptions of diffraction is that at different energies, different parts of the phase space are accessed by different experimental set-ups, as we show in the right hand plot of Fig.~\ref{fig:elastic-onech-diff}. 
 For the argument to follow, we consider
 an estimate of $\sigma_{Diff}$  given by  Eq.~(36) of \cite{Engel:2012pa},
 which provides a good interpolation  of Single Diffractive (SD)  data, from ISR to the LHC results from ALICE, CMS and  TOTEM, as we shown in Fig.~\ref{fig:elastic-onech-diff}.  i.e. 
\be
\sigma_{Diff}(s) = [\frac{(0.5 mb)\ s}{s + (10\ GeV)^2}] \log (\frac{10^3 s}{GeV^2}), \label{eq:sigdiff-EU}
\ee
We have adopted this parameterization for the full diffractive component at high energy. This is an approximation, justified at very high energy by the TOTEM result for Double Diffraction(DD) \cite{Antchev:2013any}, namely $\sigma_{DD}\simeq 0.1 mb$, although 
this result was obtained in a narrow range of pseudo rapidity and more data are needed to conclude that DD does not play a significant role at LHC energies.
 At lower energy the definitions vary, as we show in this figure. 

We shall now show how the one-channel mini-jet model presented here can be used to predict  the full inelastic cross section at higher energies.

We start with the elastic cross section, and consider  now  the difference
\be
\sigma_{elastic}^{one-ch}=\sigma_{tot}-\sigma_{inel}^{one-ch}
\ee   
which includes diffractive (otherwise said, correlated inelastic) contribution, as also discussed in general terms 
in \cite{Kopeliovich:2003tz}, among others. If
\be
\sigma_{inel}=\sigma_{inel}^{one-ch}+\sigma_{Diff}\label{eq:sigineltrue}
\ee
then, we should be able to obtain the  measured  elastic cross section from 
\be
\sigma_{elastic}=\sigma_{elastic}^{one-ch}-\sigma_{Diff} \label{eq:sigeltrue}
\ee

\begin{figure}
\resizebox{1.2\textwidth}{!}{
 \hspace{-3cm}
  \includegraphics{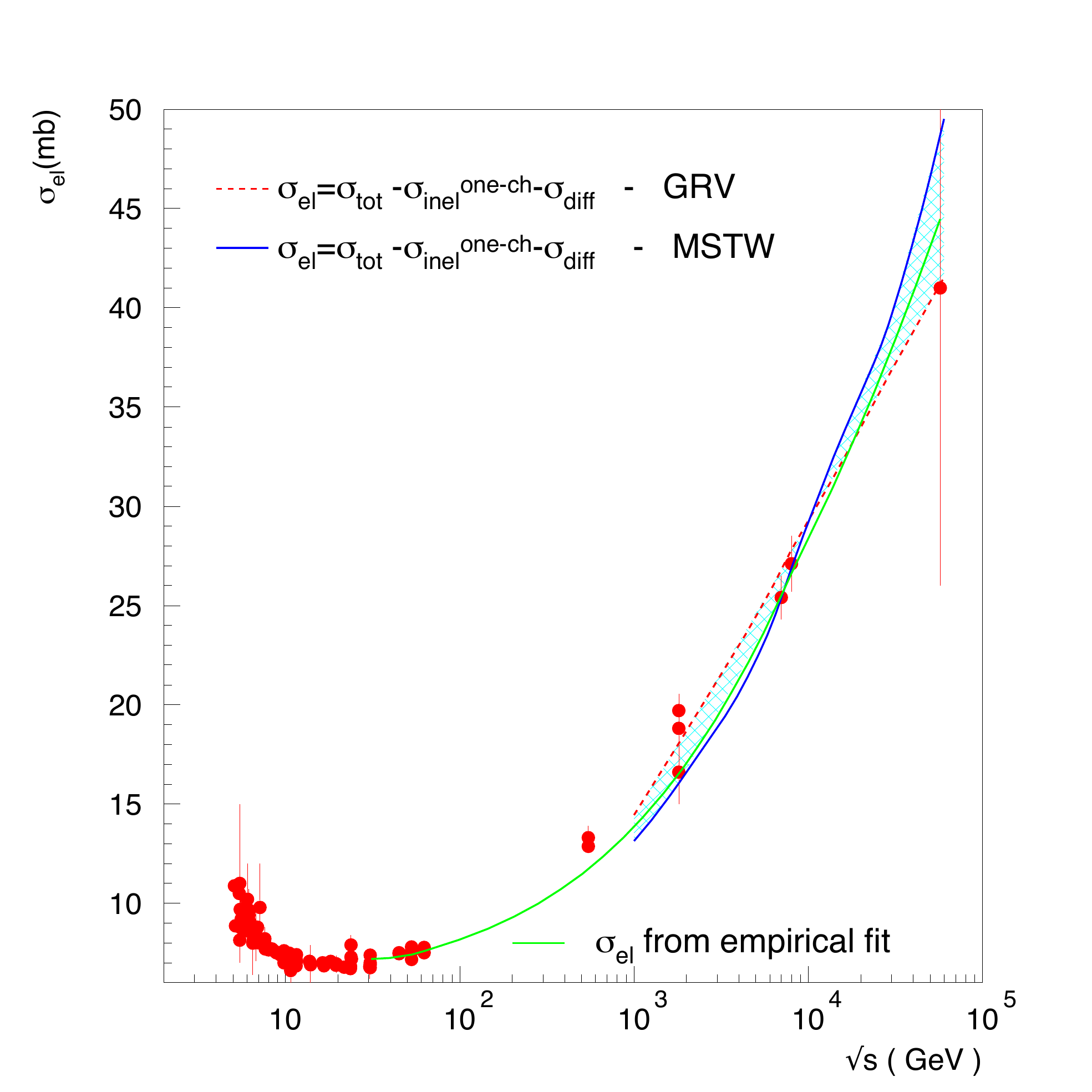}
\includegraphics{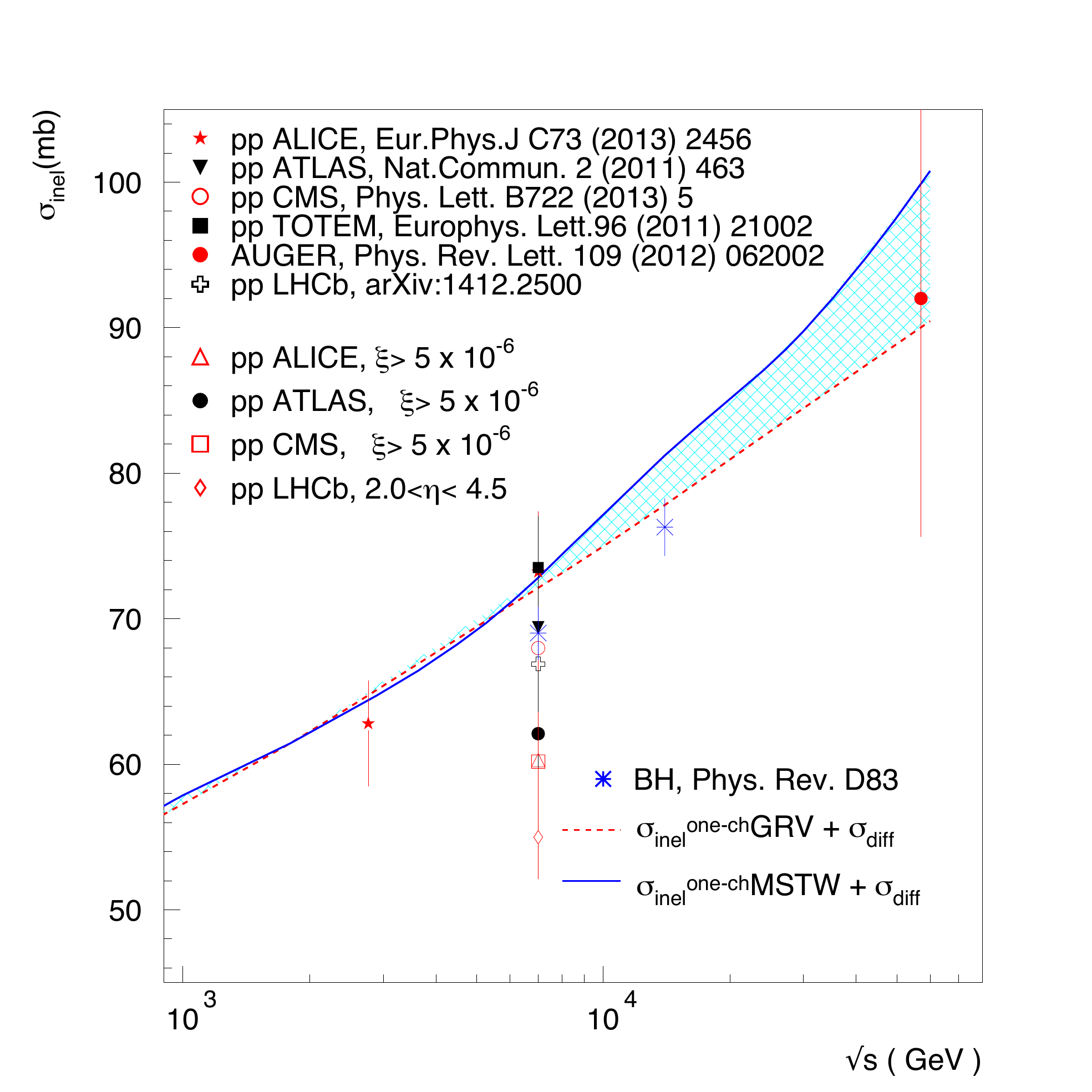}}
  \caption{Left panel: the total elastic cross section obtained by subtracting Single  Diffractive contributions,  indicated as $\sigma_{diff}$,  from the one-  channel model result.  The resulting curve is compared with $pp$ and $p {\bar p}$  data and  the empirical parametrization of \cite{Fagundes:2013aja}, which is seen to fall within the two model predictions. The right panel shows the corresponding exercise for the inelastic cross section: at high energies, adding diffraction brings the one-channel result in agreement to data.}
 \label{fig:sigeltrue}
 \end{figure}

We compare the procedure outlined through Eqs. ~(\ref{eq:sigdiff-EU}) and (\ref{eq:sigeltrue})    with experimental data and with the   empirical parametrization of Eq.~(\ref{eq:sigelemp}).  The result
is shown in the left panel of Fig.~\ref{fig:sigeltrue}. We see that 
 such a procedure gives a good description of the elastic cross section at high energy, basically past the CERN $Sp{\bar p}S$.   

 Likewise, from Eq.~(\ref{eq:sigineltrue}),  we can see that  by adding the diffractive part, parametrized as in Eq.~(\ref{eq:sigdiff-EU}), to the prediction from the one-channel model, it is possible to obtain a good description of  the high energy behavior of the inelastic cross section. This is shown in the right hand panel of Fig.~\ref{fig:sigeltrue}. It must be noticed that this procedure shows agreement with data only past ISR  energies (in fact from $Sp{\bar p}S$ onwards) energies and that a   model  describing both the low and the high energy will have to go beyond the one-channel exercise described here.  In Table \ref{tab:13tev}, we show the predictions from this model for  the inelastic cross section at LHC13, $\sqrt{s}=13\ TeV$.\footnote{
 {In the arXiv v2 and in the published PRD version, this table had an error in the MSTW predictions at $\sqrt{s}=13\ TeV$. It has now been corrected.}}

\begin{table}[H]
\centering
\caption{Minijet model predictions for the inelastic cross section  at $\sqrt{s}=13\ TeV$. Predictions of
$\sigma_{inel}$ in the full phase-space were obtained by adding  $\sigma_{diff} (13\
\text{TeV})= 12.9$ mb to  $\sigma_{inel}^{uncorr}\equiv \sigma_{inel}^{one-ch}$.\label{tab:13tev}}
\vspace*{.2cm}
\begin{tabular}{c|c|c}
\hline \hline
$PDF$ & $\sigma_{inel}^{uncorr}$ (mb) &
$\sigma_{inel}$ (mb)\\
\hline
$GRV$ &  64.3  & 77.2 \\
\hline
$MSTW$ & {66.9} &{79.8} \\
\hline \hline
\end{tabular}
\end{table}

The result of this subsection confirms the interpretation  that  at high energies, past the beginning of the 
rise and the onset of mini-jets, 
 the one-channel inelastic cross section  is  devoid of most of the diffractive contribution.
\section{Summary and Conclusions \label{Con}}
We have shown that the onset and rise of the mini-jet cross section provide the dynamical mechanism
behind the appearance of a {\it soft-edge} \cite{Block:2014lna}, i.e., a threshold in the total cross section around 
$\sqrt{s}\simeq (10\div 20)\ GeV$. Thus, our model for the total $pp$ cross section that utilizes mini-jets with 
soft-gluon re-summation has a built in soft-edge. It has been updated with recent PDFs for LHC 
at $\sqrt{s} = 7, 8\ TeV$ and predictions made for higher energy LHC data and cosmic rays. 

We have also discussed in detail the reasons behind failures to obtain correct values for the elastic cross sections 
from a one-channel eikonal that obtains the total cross section correctly. It has been shown, through the
use of phenomenological descriptions of diffractive (otherwise said, correlated inelastic) cross sections, that
one-channel elastic cross section is indeed a sum of the true elastic plus correlated inelastic cross sections.   
An application of this fact to cosmic ray data analysis for the extraction of $pp$ uncorrelated-inelastic 
cross sections shall be presented elsewhere.

 \section*{Acknowledgments}
 We thank Simone Pacetti for collaboration  about the empirical model results.
A.G. acknowledges partial support by Junta de Andalucia (FQM 6552, FQM 101). D.A.F. acknowledges the S\~ao Paulo Research Foundation (FAPESP) and CAPES
for financial support (contract: 2014/00337-8). 
O.S. acknowledges partial support from funds of Foundation of Polish Science
grant POMOST/2013-7/12, that is co-financed from European Union, Regional
Development Fund.

\bibliography{cosmic-models-2014,cosmic-data,edge}

\begin{thebibliography}{63}%
\makeatletter
\providecommand \@ifxundefined [1]{%
 \@ifx{#1\undefined}
}%
\providecommand \@ifnum [1]{%
 \ifnum #1\expandafter \@firstoftwo
 \else \expandafter \@secondoftwo
 \fi
}%
\providecommand \@ifx [1]{%
 \ifx #1\expandafter \@firstoftwo
 \else \expandafter \@secondoftwo
 \fi
}%
\providecommand \natexlab [1]{#1}%
\providecommand \enquote  [1]{``#1''}%
\providecommand \bibnamefont  [1]{#1}%
\providecommand \bibfnamefont [1]{#1}%
\providecommand \citenamefont [1]{#1}%
\providecommand \href@noop [0]{\@secondoftwo}%
\providecommand \href [0]{\begingroup \@sanitize@url \@href}%
\providecommand \@href[1]{\@@startlink{#1}\@@href}%
\providecommand \@@href[1]{\endgroup#1\@@endlink}%
\providecommand \@sanitize@url [0]{\catcode `\\12\catcode `\$12\catcode
  `\&12\catcode `\#12\catcode `\^12\catcode `\_12\catcode `\%12\relax}%
\providecommand \@@startlink[1]{}%
\providecommand \@@endlink[0]{}%
\providecommand \url  [0]{\begingroup\@sanitize@url \@url }%
\providecommand \@url [1]{\endgroup\@href {#1}{\urlprefix }}%
\providecommand \urlprefix  [0]{URL }%
\providecommand \Eprint [0]{\href }%
\providecommand \doibase [0]{http://dx.doi.org/}%
\providecommand \selectlanguage [0]{\@gobble}%
\providecommand \bibinfo  [0]{\@secondoftwo}%
\providecommand \bibfield  [0]{\@secondoftwo}%
\providecommand \translation [1]{[#1]}%
\providecommand \BibitemOpen [0]{}%
\providecommand \bibitemStop [0]{}%
\providecommand \bibitemNoStop [0]{.\EOS\space}%
\providecommand \EOS [0]{\spacefactor3000\relax}%
\providecommand \BibitemShut  [1]{\csname bibitem#1\endcsname}%
\let\auto@bib@innerbib\@empty
\bibitem [{\citenamefont {Block}\ \emph {et~al.}(2015)\citenamefont {Block},
  \citenamefont {Durand}, \citenamefont {Halzen}, \citenamefont {Stodolsky},\
  and\ \citenamefont {Weiler}}]{Block:2014lna}%
  \BibitemOpen
  \bibfield  {author} {\bibinfo {author} {\bibfnamefont {M.~M.}\ \bibnamefont
  {Block}}, \bibinfo {author} {\bibfnamefont {L.}~\bibnamefont {Durand}},
  \bibinfo {author} {\bibfnamefont {F.}~\bibnamefont {Halzen}}, \bibinfo
  {author} {\bibfnamefont {L.}~\bibnamefont {Stodolsky}}, \ and\ \bibinfo
  {author} {\bibfnamefont {T.~J.}\ \bibnamefont {Weiler}},\ }\href {\doibase
  10.1103/PhysRevD.91.011501} {\bibfield  {journal} {\bibinfo  {journal}
  {Phys.Rev.}\ }\textbf {\bibinfo {volume} {D91}},\ \bibinfo {pages} {011501}
  (\bibinfo {year} {2015})},\ \Eprint {http://arxiv.org/abs/1409.3196}
  {arXiv:1409.3196 [hep-ph]} \BibitemShut {NoStop}%
\bibitem [{\citenamefont {Achilli}\ \emph {et~al.}(2008)\citenamefont
  {Achilli}, \citenamefont {Hegde}, \citenamefont {Godbole}, \citenamefont
  {Grau}, \citenamefont {Pancheri} \emph {et~al.}}]{Achilli:2007pn}%
  \BibitemOpen
  \bibfield  {author} {\bibinfo {author} {\bibfnamefont {A.}~\bibnamefont
  {Achilli}}, \bibinfo {author} {\bibfnamefont {R.}~\bibnamefont {Hegde}},
  \bibinfo {author} {\bibfnamefont {R.~M.}\ \bibnamefont {Godbole}}, \bibinfo
  {author} {\bibfnamefont {A.}~\bibnamefont {Grau}}, \bibinfo {author}
  {\bibfnamefont {G.}~\bibnamefont {Pancheri}},  \emph {et~al.},\ }\href
  {\doibase 10.1016/j.physletb.2007.09.078} {\bibfield  {journal} {\bibinfo
  {journal} {Phys.Lett.}\ }\textbf {\bibinfo {volume} {B659}},\ \bibinfo
  {pages} {137} (\bibinfo {year} {2008})},\ \Eprint
  {http://arxiv.org/abs/0708.3626} {arXiv:0708.3626 [hep-ph]} \BibitemShut
  {NoStop}%
\bibitem [{\citenamefont {Achilli}\ \emph {et~al.}(2011)\citenamefont
  {Achilli}, \citenamefont {Godbole}, \citenamefont {Grau}, \citenamefont
  {Pancheri}, \citenamefont {Shekhovtsova} \emph {et~al.}}]{Achilli:2011sw}%
  \BibitemOpen
  \bibfield  {author} {\bibinfo {author} {\bibfnamefont {A.}~\bibnamefont
  {Achilli}}, \bibinfo {author} {\bibfnamefont {R.~M.}\ \bibnamefont
  {Godbole}}, \bibinfo {author} {\bibfnamefont {A.}~\bibnamefont {Grau}},
  \bibinfo {author} {\bibfnamefont {G.}~\bibnamefont {Pancheri}}, \bibinfo
  {author} {\bibfnamefont {O.}~\bibnamefont {Shekhovtsova}},  \emph {et~al.},\
  }\href {\doibase 10.1103/PhysRevD.84.094009} {\bibfield  {journal} {\bibinfo
  {journal} {Phys.Rev.}\ }\textbf {\bibinfo {volume} {D84}},\ \bibinfo {pages}
  {094009} (\bibinfo {year} {2011})},\ \Eprint {http://arxiv.org/abs/1102.1949}
  {arXiv:1102.1949 [hep-ph]} \BibitemShut {NoStop}%
\bibitem [{\citenamefont {Rosner}(2014)}]{Rosner:2014nka}%
  \BibitemOpen
  \bibfield  {author} {\bibinfo {author} {\bibfnamefont {J.~L.}\ \bibnamefont
  {Rosner}},\ }\href {\doibase 10.1103/PhysRevD.90.117902} {\bibfield
  {journal} {\bibinfo  {journal} {Phys.Rev.}\ }\textbf {\bibinfo {volume}
  {D90}},\ \bibinfo {pages} {117902} (\bibinfo {year} {2014})},\ \Eprint
  {http://arxiv.org/abs/1409.5813} {arXiv:1409.5813 [hep-ph]} \BibitemShut
  {NoStop}%
\bibitem [{\citenamefont {Grau}\ \emph {et~al.}(1999)\citenamefont {Grau},
  \citenamefont {Pancheri},\ and\ \citenamefont {Srivastava}}]{Grau:1999em}%
  \BibitemOpen
  \bibfield  {author} {\bibinfo {author} {\bibfnamefont {A.}~\bibnamefont
  {Grau}}, \bibinfo {author} {\bibfnamefont {G.}~\bibnamefont {Pancheri}}, \
  and\ \bibinfo {author} {\bibfnamefont {Y.}~\bibnamefont {Srivastava}},\
  }\href {\doibase 10.1103/PhysRevD.60.114020} {\bibfield  {journal} {\bibinfo
  {journal} {Phys.Rev.}\ }\textbf {\bibinfo {volume} {D60}},\ \bibinfo {pages}
  {114020} (\bibinfo {year} {1999})},\ \Eprint
  {http://arxiv.org/abs/hep-ph/9905228} {arXiv:hep-ph/9905228 [hep-ph]}
  \BibitemShut {NoStop}%
\bibitem [{\citenamefont {Gotsman}\ \emph
  {et~al.}(2012{\natexlab{a}})\citenamefont {Gotsman}, \citenamefont {Levin},\
  and\ \citenamefont {Maor}}]{Gotsman:2012rm}%
  \BibitemOpen
  \bibfield  {author} {\bibinfo {author} {\bibfnamefont {E.}~\bibnamefont
  {Gotsman}}, \bibinfo {author} {\bibfnamefont {E.}~\bibnamefont {Levin}}, \
  and\ \bibinfo {author} {\bibfnamefont {U.}~\bibnamefont {Maor}},\ }\href
  {\doibase 10.1016/j.physletb.2012.08.042} {\bibfield  {journal} {\bibinfo
  {journal} {Phys.Lett.}\ }\textbf {\bibinfo {volume} {B716}},\ \bibinfo
  {pages} {425} (\bibinfo {year} {2012}{\natexlab{a}})},\ \Eprint
  {http://arxiv.org/abs/1208.0898} {arXiv:1208.0898 [hep-ph]} \BibitemShut
  {NoStop}%
\bibitem [{\citenamefont {Khoze}\ \emph {et~al.}(2015)\citenamefont {Khoze},
  \citenamefont {Martin},\ and\ \citenamefont {Ryskin}}]{Khoze:2014nia}%
  \BibitemOpen
  \bibfield  {author} {\bibinfo {author} {\bibfnamefont {V.}~\bibnamefont
  {Khoze}}, \bibinfo {author} {\bibfnamefont {A.}~\bibnamefont {Martin}}, \
  and\ \bibinfo {author} {\bibfnamefont {M.}~\bibnamefont {Ryskin}},\ }\href
  {\doibase 10.1088/0954-3899/42/2/025003} {\bibfield  {journal} {\bibinfo
  {journal} {J.Phys.}\ }\textbf {\bibinfo {volume} {G42}},\ \bibinfo {pages}
  {025003} (\bibinfo {year} {2015})},\ \Eprint {http://arxiv.org/abs/1410.0508}
  {arXiv:1410.0508 [hep-ph]} \BibitemShut {NoStop}%
\bibitem [{\citenamefont {Gluck}\ \emph {et~al.}(1992)\citenamefont {Gluck},
  \citenamefont {Reya},\ and\ \citenamefont {Vogt}}]{Gluck:1991ng}%
  \BibitemOpen
  \bibfield  {author} {\bibinfo {author} {\bibfnamefont {M.}~\bibnamefont
  {Gluck}}, \bibinfo {author} {\bibfnamefont {E.}~\bibnamefont {Reya}}, \ and\
  \bibinfo {author} {\bibfnamefont {A.}~\bibnamefont {Vogt}},\ }\href {\doibase
  10.1007/BF01483880} {\bibfield  {journal} {\bibinfo  {journal} {Z. Phys.}\
  }\textbf {\bibinfo {volume} {C53}},\ \bibinfo {pages} {127} (\bibinfo {year}
  {1992})}\BibitemShut {NoStop}%
\bibitem [{\citenamefont {Gluck}\ \emph {et~al.}(1995)\citenamefont {Gluck},
  \citenamefont {Reya},\ and\ \citenamefont {Vogt}}]{Gluck:1994uf}%
  \BibitemOpen
  \bibfield  {author} {\bibinfo {author} {\bibfnamefont {M.}~\bibnamefont
  {Gluck}}, \bibinfo {author} {\bibfnamefont {E.}~\bibnamefont {Reya}}, \ and\
  \bibinfo {author} {\bibfnamefont {A.}~\bibnamefont {Vogt}},\ }\href {\doibase
  10.1007/BF01624586} {\bibfield  {journal} {\bibinfo  {journal} {Z. Phys.}\
  }\textbf {\bibinfo {volume} {C67}},\ \bibinfo {pages} {433} (\bibinfo {year}
  {1995})}\BibitemShut {NoStop}%
\bibitem [{\citenamefont {Gluck}\ \emph {et~al.}(1998)\citenamefont {Gluck},
  \citenamefont {Reya},\ and\ \citenamefont {Vogt}}]{Gluck:1998xa}%
  \BibitemOpen
  \bibfield  {author} {\bibinfo {author} {\bibfnamefont {M.}~\bibnamefont
  {Gluck}}, \bibinfo {author} {\bibfnamefont {E.}~\bibnamefont {Reya}}, \ and\
  \bibinfo {author} {\bibfnamefont {A.}~\bibnamefont {Vogt}},\ }\href {\doibase
  10.1007/s100520050289} {\bibfield  {journal} {\bibinfo  {journal} {Eur. Phys.
  J.}\ }\textbf {\bibinfo {volume} {C5}},\ \bibinfo {pages} {461} (\bibinfo
  {year} {1998})},\ \Eprint {http://arxiv.org/abs/hep-ph/9806404}
  {arXiv:hep-ph/9806404} \BibitemShut {NoStop}%
\bibitem [{\citenamefont {Martin}\ \emph {et~al.}(1998)\citenamefont {Martin},
  \citenamefont {Roberts}, \citenamefont {Stirling},\ and\ \citenamefont
  {Thorne}}]{Martin:1998sq}%
  \BibitemOpen
  \bibfield  {author} {\bibinfo {author} {\bibfnamefont {A.~D.}\ \bibnamefont
  {Martin}}, \bibinfo {author} {\bibfnamefont {R.~G.}\ \bibnamefont {Roberts}},
  \bibinfo {author} {\bibfnamefont {W.~J.}\ \bibnamefont {Stirling}}, \ and\
  \bibinfo {author} {\bibfnamefont {R.~S.}\ \bibnamefont {Thorne}},\ }\href
  {\doibase 10.1007/s100520050220} {\bibfield  {journal} {\bibinfo  {journal}
  {Eur. Phys. J.}\ }\textbf {\bibinfo {volume} {C4}},\ \bibinfo {pages} {463}
  (\bibinfo {year} {1998})},\ \Eprint {http://arxiv.org/abs/hep-ph/9803445}
  {arXiv:hep-ph/9803445} \BibitemShut {NoStop}%
\bibitem [{\citenamefont {Martin}\ \emph {et~al.}(2009)\citenamefont {Martin},
  \citenamefont {Stirling}, \citenamefont {Thorne},\ and\ \citenamefont
  {Watt}}]{Martin:2009iq}%
  \BibitemOpen
  \bibfield  {author} {\bibinfo {author} {\bibfnamefont {A.}~\bibnamefont
  {Martin}}, \bibinfo {author} {\bibfnamefont {W.}~\bibnamefont {Stirling}},
  \bibinfo {author} {\bibfnamefont {R.}~\bibnamefont {Thorne}}, \ and\ \bibinfo
  {author} {\bibfnamefont {G.}~\bibnamefont {Watt}},\ }\href {\doibase
  10.1140/epjc/s10052-009-1072-5} {\bibfield  {journal} {\bibinfo  {journal}
  {Eur.Phys.J.}\ }\textbf {\bibinfo {volume} {C63}},\ \bibinfo {pages} {189}
  (\bibinfo {year} {2009})},\ \Eprint {http://arxiv.org/abs/0901.0002}
  {arXiv:0901.0002 [hep-ph]} \BibitemShut {NoStop}%
\bibitem [{\citenamefont {Godbole}\ \emph {et~al.}(2005)\citenamefont
  {Godbole}, \citenamefont {Grau}, \citenamefont {Pancheri},\ and\
  \citenamefont {Srivastava}}]{Godbole:2004kx}%
  \BibitemOpen
  \bibfield  {author} {\bibinfo {author} {\bibfnamefont {R.~M.}\ \bibnamefont
  {Godbole}}, \bibinfo {author} {\bibfnamefont {A.}~\bibnamefont {Grau}},
  \bibinfo {author} {\bibfnamefont {G.}~\bibnamefont {Pancheri}}, \ and\
  \bibinfo {author} {\bibfnamefont {Y.~N.}\ \bibnamefont {Srivastava}},\ }\href
  {\doibase 10.1103/PhysRevD.72.076001} {\bibfield  {journal} {\bibinfo
  {journal} {Phys. Rev.}\ }\textbf {\bibinfo {volume} {D72}},\ \bibinfo {pages}
  {076001} (\bibinfo {year} {2005})},\ \Eprint
  {http://arxiv.org/abs/hep-ph/0408355} {arXiv:hep-ph/0408355} \BibitemShut
  {NoStop}%
\bibitem [{\citenamefont {Corsetti}\ \emph {et~al.}(1996)\citenamefont
  {Corsetti}, \citenamefont {Grau}, \citenamefont {Pancheri},\ and\
  \citenamefont {Srivastava}}]{Corsetti:1996wg}%
  \BibitemOpen
  \bibfield  {author} {\bibinfo {author} {\bibfnamefont {A.}~\bibnamefont
  {Corsetti}}, \bibinfo {author} {\bibfnamefont {A.}~\bibnamefont {Grau}},
  \bibinfo {author} {\bibfnamefont {G.}~\bibnamefont {Pancheri}}, \ and\
  \bibinfo {author} {\bibfnamefont {Y.~N.}\ \bibnamefont {Srivastava}},\ }\href
  {\doibase 10.1016/0370-2693(96)00566-7} {\bibfield  {journal} {\bibinfo
  {journal} {Phys. Lett.}\ }\textbf {\bibinfo {volume} {B382}},\ \bibinfo
  {pages} {282} (\bibinfo {year} {1996})},\ \Eprint
  {http://arxiv.org/abs/hep-ph/9605314} {arXiv:hep-ph/9605314} \BibitemShut
  {NoStop}%
\bibitem [{\citenamefont {Dokshitzer}\ \emph {et~al.}(1978)\citenamefont
  {Dokshitzer}, \citenamefont {Diakonov},\ and\ \citenamefont
  {Troian}}]{Dokshitzer:1978yd}%
  \BibitemOpen
  \bibfield  {author} {\bibinfo {author} {\bibfnamefont {Y.~L.}\ \bibnamefont
  {Dokshitzer}}, \bibinfo {author} {\bibfnamefont {D.}~\bibnamefont
  {Diakonov}}, \ and\ \bibinfo {author} {\bibfnamefont {S.~I.}\ \bibnamefont
  {Troian}},\ }\href {\doibase 10.1016/0370-2693(78)90240-X} {\bibfield
  {journal} {\bibinfo  {journal} {Phys. Lett.}\ }\textbf {\bibinfo {volume}
  {B79}},\ \bibinfo {pages} {269} (\bibinfo {year} {1978})}\BibitemShut
  {NoStop}%
\bibitem [{\citenamefont {Parisi}\ and\ \citenamefont
  {Petronzio}(1979)}]{Parisi:1979se}%
  \BibitemOpen
  \bibfield  {author} {\bibinfo {author} {\bibfnamefont {G.}~\bibnamefont
  {Parisi}}\ and\ \bibinfo {author} {\bibfnamefont {R.}~\bibnamefont
  {Petronzio}},\ }\href {\doibase 10.1016/0550-3213(79)90040-3} {\bibfield
  {journal} {\bibinfo  {journal} {Nucl. Phys.}\ }\textbf {\bibinfo {volume}
  {B154}},\ \bibinfo {pages} {427} (\bibinfo {year} {1979})}\BibitemShut
  {NoStop}%
\bibitem [{\citenamefont {Bloch}\ and\ \citenamefont
  {Nordsieck}(1937)}]{Bloch:1937pw}%
  \BibitemOpen
  \bibfield  {author} {\bibinfo {author} {\bibfnamefont {F.}~\bibnamefont
  {Bloch}}\ and\ \bibinfo {author} {\bibfnamefont {A.}~\bibnamefont
  {Nordsieck}},\ }\href {\doibase 10.1103/PhysRev.52.54} {\bibfield  {journal}
  {\bibinfo  {journal} {Phys. Rev.}\ }\textbf {\bibinfo {volume} {52}},\
  \bibinfo {pages} {54} (\bibinfo {year} {1937})}\BibitemShut {NoStop}%
\bibitem [{\citenamefont {Grau}\ \emph {et~al.}(2009)\citenamefont {Grau},
  \citenamefont {Godbole}, \citenamefont {Pancheri},\ and\ \citenamefont
  {Srivastava}}]{Grau:2009qx}%
  \BibitemOpen
  \bibfield  {author} {\bibinfo {author} {\bibfnamefont {A.}~\bibnamefont
  {Grau}}, \bibinfo {author} {\bibfnamefont {R.~M.}\ \bibnamefont {Godbole}},
  \bibinfo {author} {\bibfnamefont {G.}~\bibnamefont {Pancheri}}, \ and\
  \bibinfo {author} {\bibfnamefont {Y.~N.}\ \bibnamefont {Srivastava}},\ }\href
  {\doibase 10.1016/j.physletb.2009.10.080} {\bibfield  {journal} {\bibinfo
  {journal} {Phys.Lett.}\ }\textbf {\bibinfo {volume} {B682}},\ \bibinfo
  {pages} {55} (\bibinfo {year} {2009})},\ \Eprint
  {http://arxiv.org/abs/0908.1426} {arXiv:0908.1426 [hep-ph]} \BibitemShut
  {NoStop}%
\bibitem [{\citenamefont {Chiappetta}\ and\ \citenamefont
  {Greco}(1981)}]{Chiappetta:1981bw}%
  \BibitemOpen
  \bibfield  {author} {\bibinfo {author} {\bibfnamefont {P.}~\bibnamefont
  {Chiappetta}}\ and\ \bibinfo {author} {\bibfnamefont {M.}~\bibnamefont
  {Greco}},\ }\href {\doibase 10.1016/0370-2693(81)90912-6} {\bibfield
  {journal} {\bibinfo  {journal} {Phys. Lett.}\ }\textbf {\bibinfo {volume}
  {B106}},\ \bibinfo {pages} {219} (\bibinfo {year} {1981})}\BibitemShut
  {NoStop}%
\bibitem [{\citenamefont {Pancheri}\ and\ \citenamefont
  {Srivastava}(1985)}]{Pancheri:1985ix}%
  \BibitemOpen
  \bibfield  {author} {\bibinfo {author} {\bibfnamefont {G.}~\bibnamefont
  {Pancheri}}\ and\ \bibinfo {author} {\bibfnamefont {Y.}~\bibnamefont
  {Srivastava}},\ }\href {\doibase 10.1016/0370-2693(85)90121-2} {\bibfield
  {journal} {\bibinfo  {journal} {Phys.Lett.}\ }\textbf {\bibinfo {volume}
  {B159}},\ \bibinfo {pages} {69} (\bibinfo {year} {1985})}\BibitemShut
  {NoStop}%
\bibitem [{\citenamefont {Gaisser}\ and\ \citenamefont
  {Halzen}(1985)}]{Gaisser:1984pg}%
  \BibitemOpen
  \bibfield  {author} {\bibinfo {author} {\bibfnamefont {T.~K.}\ \bibnamefont
  {Gaisser}}\ and\ \bibinfo {author} {\bibfnamefont {F.}~\bibnamefont
  {Halzen}},\ }\href {\doibase 10.1103/PhysRevLett.54.1754} {\bibfield
  {journal} {\bibinfo  {journal} {Phys. Rev. Lett.}\ }\textbf {\bibinfo
  {volume} {54}},\ \bibinfo {pages} {1754} (\bibinfo {year}
  {1985})}\BibitemShut {NoStop}%
\bibitem [{\citenamefont {Durand}\ and\ \citenamefont
  {Pi}(1989)}]{Durand:1988ax}%
  \BibitemOpen
  \bibfield  {author} {\bibinfo {author} {\bibfnamefont {L.}~\bibnamefont
  {Durand}}\ and\ \bibinfo {author} {\bibfnamefont {H.}~\bibnamefont {Pi}},\
  }\href {\doibase 10.1103/PhysRevD.40.1436} {\bibfield  {journal} {\bibinfo
  {journal} {Phys. Rev.}\ }\textbf {\bibinfo {volume} {D40}},\ \bibinfo {pages}
  {1436} (\bibinfo {year} {1989})}\BibitemShut {NoStop}%
\bibitem [{\citenamefont {Khoze}\ \emph {et~al.}(2014)\citenamefont {Khoze},
  \citenamefont {Martin},\ and\ \citenamefont {Ryskin}}]{Khoze:2013jsa}%
  \BibitemOpen
  \bibfield  {author} {\bibinfo {author} {\bibfnamefont {V.}~\bibnamefont
  {Khoze}}, \bibinfo {author} {\bibfnamefont {A.}~\bibnamefont {Martin}}, \
  and\ \bibinfo {author} {\bibfnamefont {M.}~\bibnamefont {Ryskin}},\ }\href
  {\doibase 10.1140/epjc/s10052-014-2756-z} {\bibfield  {journal} {\bibinfo
  {journal} {Eur.Phys.J.}\ }\textbf {\bibinfo {volume} {C74}},\ \bibinfo
  {pages} {2756} (\bibinfo {year} {2014})},\ \Eprint
  {http://arxiv.org/abs/1312.3851} {arXiv:1312.3851 [hep-ph]} \BibitemShut
  {NoStop}%
\bibitem [{\citenamefont {Good}\ and\ \citenamefont
  {Walker}(1960)}]{Good:1960ba}%
  \BibitemOpen
  \bibfield  {author} {\bibinfo {author} {\bibfnamefont {M.}~\bibnamefont
  {Good}}\ and\ \bibinfo {author} {\bibfnamefont {W.}~\bibnamefont {Walker}},\
  }\href {\doibase 10.1103/PhysRev.120.1857} {\bibfield  {journal} {\bibinfo
  {journal} {Phys.Rev.}\ }\textbf {\bibinfo {volume} {120}},\ \bibinfo {pages}
  {1857} (\bibinfo {year} {1960})}\BibitemShut {NoStop}%
\bibitem [{\citenamefont {Lipari}\ and\ \citenamefont
  {Lusignoli}(2009)}]{Lipari:2009rm}%
  \BibitemOpen
  \bibfield  {author} {\bibinfo {author} {\bibfnamefont {P.}~\bibnamefont
  {Lipari}}\ and\ \bibinfo {author} {\bibfnamefont {M.}~\bibnamefont
  {Lusignoli}},\ }\href {\doibase 10.1103/PhysRevD.80.074014} {\bibfield
  {journal} {\bibinfo  {journal} {Phys. Rev.}\ }\textbf {\bibinfo {volume}
  {D80}},\ \bibinfo {pages} {074014} (\bibinfo {year} {2009})},\ \Eprint
  {http://arxiv.org/abs/0908.0495} {arXiv:0908.0495 [hep-ph]} \BibitemShut
  {NoStop}%
\bibitem [{\citenamefont {Lipari}\ and\ \citenamefont
  {Lusignoli}(2013)}]{Lipari:2013kta}%
  \BibitemOpen
  \bibfield  {author} {\bibinfo {author} {\bibfnamefont {P.}~\bibnamefont
  {Lipari}}\ and\ \bibinfo {author} {\bibfnamefont {M.}~\bibnamefont
  {Lusignoli}},\ }\href {\doibase 10.1140/epjc/s10052-013-2630-4} {\bibfield
  {journal} {\bibinfo  {journal} {Eur.Phys.J.}\ }\textbf {\bibinfo {volume}
  {C73}},\ \bibinfo {pages} {2630} (\bibinfo {year} {2013})},\ \Eprint
  {http://arxiv.org/abs/1305.7216} {arXiv:1305.7216 [hep-ph]} \BibitemShut
  {NoStop}%
\bibitem [{\citenamefont {Ryskin}\ \emph {et~al.}(2012)\citenamefont {Ryskin},
  \citenamefont {Martin},\ and\ \citenamefont {Khoze}}]{Ryskin:2012az}%
  \BibitemOpen
  \bibfield  {author} {\bibinfo {author} {\bibfnamefont {M.}~\bibnamefont
  {Ryskin}}, \bibinfo {author} {\bibfnamefont {A.}~\bibnamefont {Martin}}, \
  and\ \bibinfo {author} {\bibfnamefont {V.}~\bibnamefont {Khoze}},\ }\href
  {\doibase 10.1140/epjc/s10052-012-1937-x} {\bibfield  {journal} {\bibinfo
  {journal} {Eur.Phys.J.}\ }\textbf {\bibinfo {volume} {C72}},\ \bibinfo
  {pages} {1937} (\bibinfo {year} {2012})},\ \Eprint
  {http://arxiv.org/abs/1201.6298} {arXiv:1201.6298 [hep-ph]} \BibitemShut
  {NoStop}%
\bibitem [{\citenamefont {Gotsman}\ \emph
  {et~al.}(2012{\natexlab{b}})\citenamefont {Gotsman}, \citenamefont {Levin},\
  and\ \citenamefont {Maor}}]{Gotsman:2012rq}%
  \BibitemOpen
  \bibfield  {author} {\bibinfo {author} {\bibfnamefont {E.}~\bibnamefont
  {Gotsman}}, \bibinfo {author} {\bibfnamefont {E.}~\bibnamefont {Levin}}, \
  and\ \bibinfo {author} {\bibfnamefont {U.}~\bibnamefont {Maor}},\ }\href
  {\doibase 10.1103/PhysRevD.85.094007} {\bibfield  {journal} {\bibinfo
  {journal} {Phys.Rev.}\ }\textbf {\bibinfo {volume} {D85}},\ \bibinfo {pages}
  {094007} (\bibinfo {year} {2012}{\natexlab{b}})},\ \Eprint
  {http://arxiv.org/abs/1203.2419} {arXiv:1203.2419 [hep-ph]} \BibitemShut
  {NoStop}%
\bibitem [{\citenamefont {Ostapchenko}(2011)}]{Ostapchenko:2010vb}%
  \BibitemOpen
  \bibfield  {author} {\bibinfo {author} {\bibfnamefont {S.}~\bibnamefont
  {Ostapchenko}},\ }\href {\doibase 10.1103/PhysRevD.83.014018} {\bibfield
  {journal} {\bibinfo  {journal} {Phys.Rev.}\ }\textbf {\bibinfo {volume}
  {D83}},\ \bibinfo {pages} {014018} (\bibinfo {year} {2011})},\ \Eprint
  {http://arxiv.org/abs/1010.1869} {arXiv:1010.1869 [hep-ph]} \BibitemShut
  {NoStop}%
\bibitem [{\citenamefont {Balitsky}(1996)}]{Balitsky:1995ub}%
  \BibitemOpen
  \bibfield  {author} {\bibinfo {author} {\bibfnamefont {I.}~\bibnamefont
  {Balitsky}},\ }\href {\doibase 10.1016/0550-3213(95)00638-9} {\bibfield
  {journal} {\bibinfo  {journal} {Nucl.Phys.}\ }\textbf {\bibinfo {volume}
  {B463}},\ \bibinfo {pages} {99} (\bibinfo {year} {1996})},\ \Eprint
  {http://arxiv.org/abs/hep-ph/9509348} {arXiv:hep-ph/9509348 [hep-ph]}
  \BibitemShut {NoStop}%
\bibitem [{\citenamefont {Kovchegov}(2001)}]{Kovchegov:2001ni}%
  \BibitemOpen
  \bibfield  {author} {\bibinfo {author} {\bibfnamefont {Y.~V.}\ \bibnamefont
  {Kovchegov}},\ }\href {\doibase 10.1103/PhysRevD.64.114016,
  10.1103/PhysRevD.68.039901} {\bibfield  {journal} {\bibinfo  {journal}
  {Phys.Rev.}\ }\textbf {\bibinfo {volume} {D64}},\ \bibinfo {pages} {114016}
  (\bibinfo {year} {2001})},\ \Eprint {http://arxiv.org/abs/hep-ph/0107256}
  {arXiv:hep-ph/0107256 [hep-ph]} \BibitemShut {NoStop}%
\bibitem [{\citenamefont {Ostapchenko}(2010)}]{Ostapchenko:2010gt}%
  \BibitemOpen
  \bibfield  {author} {\bibinfo {author} {\bibfnamefont {S.}~\bibnamefont
  {Ostapchenko}},\ }\href {\doibase 10.1103/PhysRevD.81.114028} {\bibfield
  {journal} {\bibinfo  {journal} {Phys.Rev.}\ }\textbf {\bibinfo {volume}
  {D81}},\ \bibinfo {pages} {114028} (\bibinfo {year} {2010})},\ \Eprint
  {http://arxiv.org/abs/1003.0196} {arXiv:1003.0196 [hep-ph]} \BibitemShut
  {NoStop}%
\bibitem [{\citenamefont {Grau}\ \emph {et~al.}(2010)\citenamefont {Grau},
  \citenamefont {Pancheri}, \citenamefont {Shekhovtsova},\ and\ \citenamefont
  {Srivastava}}]{Grau:2010ju}%
  \BibitemOpen
  \bibfield  {author} {\bibinfo {author} {\bibfnamefont {A.}~\bibnamefont
  {Grau}}, \bibinfo {author} {\bibfnamefont {G.}~\bibnamefont {Pancheri}},
  \bibinfo {author} {\bibfnamefont {O.}~\bibnamefont {Shekhovtsova}}, \ and\
  \bibinfo {author} {\bibfnamefont {Y.~N.}\ \bibnamefont {Srivastava}},\ }\href
  {\doibase 10.1016/j.physletb.2010.08.078} {\bibfield  {journal} {\bibinfo
  {journal} {Phys.Lett.}\ }\textbf {\bibinfo {volume} {B693}},\ \bibinfo
  {pages} {456} (\bibinfo {year} {2010})},\ \Eprint
  {http://arxiv.org/abs/1008.4119} {arXiv:1008.4119 [hep-ph]} \BibitemShut
  {NoStop}%
\bibitem [{\citenamefont {Antchev}\ \emph
  {et~al.}(2013{\natexlab{a}})\citenamefont {Antchev} \emph
  {et~al.}}]{Antchev:2013paa}%
  \BibitemOpen
  \bibfield  {author} {\bibinfo {author} {\bibfnamefont {G.}~\bibnamefont
  {Antchev}} \emph {et~al.} (\bibinfo {collaboration} {TOTEM Collaboration}),\
  }\href {\doibase 10.1103/PhysRevLett.111.012001} {\bibfield  {journal}
  {\bibinfo  {journal} {Phys.Rev.Lett.}\ }\textbf {\bibinfo {volume} {111}},\
  \bibinfo {pages} {012001} (\bibinfo {year} {2013}{\natexlab{a}})}\BibitemShut
  {NoStop}%
\bibitem [{\citenamefont {Antchev}\ \emph
  {et~al.}(2013{\natexlab{b}})\citenamefont {Antchev} \emph
  {et~al.}}]{Antchev:2013iaa}%
  \BibitemOpen
  \bibfield  {author} {\bibinfo {author} {\bibfnamefont {G.}~\bibnamefont
  {Antchev}} \emph {et~al.} (\bibinfo {collaboration} {TOTEM}),\ }\href
  {\doibase 10.1209/0295-5075/101/21004} {\bibfield  {journal} {\bibinfo
  {journal} {Europhys.Lett.}\ }\textbf {\bibinfo {volume} {101}},\ \bibinfo
  {pages} {21004} (\bibinfo {year} {2013}{\natexlab{b}})}\BibitemShut {NoStop}%
\bibitem [{\citenamefont {Aad}\ \emph {et~al.}(2014)\citenamefont {Aad} \emph
  {et~al.}}]{Aad:2014dca}%
  \BibitemOpen
  \bibfield  {author} {\bibinfo {author} {\bibfnamefont {G.}~\bibnamefont
  {Aad}} \emph {et~al.} (\bibinfo {collaboration} {ATLAS}),\ }\href {\doibase
  10.1016/j.nuclphysb.2014.10.019} {\bibfield  {journal} {\bibinfo  {journal}
  {Nucl.Phys.}\ }\textbf {\bibinfo {volume} {B889}},\ \bibinfo {pages} {486}
  (\bibinfo {year} {2014})},\ \Eprint {http://arxiv.org/abs/1408.5778}
  {arXiv:1408.5778 [hep-ex]} \BibitemShut {NoStop}%
\bibitem [{\citenamefont {Block}\ and\ \citenamefont
  {Halzen}(2011{\natexlab{a}})}]{Block:2011vz}%
  \BibitemOpen
  \bibfield  {author} {\bibinfo {author} {\bibfnamefont {M.~M.}\ \bibnamefont
  {Block}}\ and\ \bibinfo {author} {\bibfnamefont {F.}~\bibnamefont {Halzen}},\
  }\href@noop {} {\bibfield  {journal} {\bibinfo  {journal} {Phys.Rev.Lett.}\
  }\textbf {\bibinfo {volume} {107}},\ \bibinfo {pages} {212002} (\bibinfo
  {year} {2011}{\natexlab{a}})},\ \Eprint {http://arxiv.org/abs/1109.2041}
  {arXiv:1109.2041 [hep-ph]} \BibitemShut {NoStop}%
\bibitem [{\citenamefont {Fagundes}\ \emph
  {et~al.}(2013{\natexlab{a}})\citenamefont {Fagundes}, \citenamefont {Menon},\
  and\ \citenamefont {Silva}}]{Fagundes:2012rr}%
  \BibitemOpen
  \bibfield  {author} {\bibinfo {author} {\bibfnamefont {D.~A.}\ \bibnamefont
  {Fagundes}}, \bibinfo {author} {\bibfnamefont {M.~J.}\ \bibnamefont {Menon}},
  \ and\ \bibinfo {author} {\bibfnamefont {P.~V. R.~G.}\ \bibnamefont
  {Silva}},\ }\href {\doibase 10.1088/0954-3899/40/6/065005} {\bibfield
  {journal} {\bibinfo  {journal} {J.Phys.}\ }\textbf {\bibinfo {volume}
  {G40}},\ \bibinfo {pages} {065005} (\bibinfo {year} {2013}{\natexlab{a}})},\
  \Eprint {http://arxiv.org/abs/1208.3456} {arXiv:1208.3456 [hep-ph]}
  \BibitemShut {NoStop}%
\bibitem [{\citenamefont {Fagundes}\ \emph {et~al.}(2015)\citenamefont
  {Fagundes}, \citenamefont {Grau}, \citenamefont {Pancheri}, \citenamefont
  {Srivastava},\ and\ \citenamefont {Shekhovtsova}}]{Fagundes:2014fza}%
  \BibitemOpen
  \bibfield  {author} {\bibinfo {author} {\bibfnamefont {D.~A.}\ \bibnamefont
  {Fagundes}}, \bibinfo {author} {\bibfnamefont {A.}~\bibnamefont {Grau}},
  \bibinfo {author} {\bibfnamefont {G.}~\bibnamefont {Pancheri}}, \bibinfo
  {author} {\bibfnamefont {Y.~N.}\ \bibnamefont {Srivastava}}, \ and\ \bibinfo
  {author} {\bibfnamefont {O.}~\bibnamefont {Shekhovtsova}},\ }\href {\doibase
  10.1051/epjconf/20159003002} {\bibfield  {journal} {\bibinfo  {journal} {EPJ
  Web Conf.}\ }\textbf {\bibinfo {volume} {90}},\ \bibinfo {pages} {03002}
  (\bibinfo {year} {2015})},\ \Eprint {http://arxiv.org/abs/1408.2921}
  {arXiv:1408.2921 [hep-ph]} \BibitemShut {NoStop}%
\bibitem [{\citenamefont {Giannini}\ and\ \citenamefont
  {Duraes}(2013)}]{Giannini:2013jla}%
  \BibitemOpen
  \bibfield  {author} {\bibinfo {author} {\bibfnamefont {A.}~\bibnamefont
  {Giannini}}\ and\ \bibinfo {author} {\bibfnamefont {F.}~\bibnamefont
  {Duraes}},\ }\href {\doibase 10.1103/PhysRevD.88.114004} {\bibfield
  {journal} {\bibinfo  {journal} {Phys.Rev.}\ }\textbf {\bibinfo {volume}
  {D88}},\ \bibinfo {pages} {114004} (\bibinfo {year} {2013})},\ \Eprint
  {http://arxiv.org/abs/1302.3765} {arXiv:1302.3765 [hep-ph]} \BibitemShut
  {NoStop}%
\bibitem [{\citenamefont {Aad}\ \emph {et~al.}(2011)\citenamefont {Aad} \emph
  {et~al.}}]{Aad:2011eu}%
  \BibitemOpen
  \bibfield  {author} {\bibinfo {author} {\bibfnamefont {G.}~\bibnamefont
  {Aad}} \emph {et~al.} (\bibinfo {collaboration} {ATLAS Collaboration}),\
  }\href {\doibase 10.1038/ncomms1472} {\bibfield  {journal} {\bibinfo
  {journal} {Nature Commun.}\ }\textbf {\bibinfo {volume} {2}},\ \bibinfo
  {pages} {463} (\bibinfo {year} {2011})},\ \Eprint
  {http://arxiv.org/abs/1104.0326} {arXiv:1104.0326 [hep-ex]} \BibitemShut
  {NoStop}%
\bibitem [{\citenamefont {Chatrchyan}\ \emph {et~al.}(2013)\citenamefont
  {Chatrchyan} \emph {et~al.}}]{Chatrchyan:2012nj}%
  \BibitemOpen
  \bibfield  {author} {\bibinfo {author} {\bibfnamefont {S.}~\bibnamefont
  {Chatrchyan}} \emph {et~al.} (\bibinfo {collaboration} {CMS Collaboration}),\
  }\href {\doibase 10.1016/j.physletb.2013.03.024} {\bibfield  {journal}
  {\bibinfo  {journal} {Phys.Lett.}\ }\textbf {\bibinfo {volume} {B722}},\
  \bibinfo {pages} {5} (\bibinfo {year} {2013})},\ \Eprint
  {http://arxiv.org/abs/1210.6718} {arXiv:1210.6718 [hep-ex]} \BibitemShut
  {NoStop}%
\bibitem [{\citenamefont {Antchev}\ \emph
  {et~al.}(2013{\natexlab{c}})\citenamefont {Antchev} \emph
  {et~al.}}]{Antchev:2013haa}%
  \BibitemOpen
  \bibfield  {author} {\bibinfo {author} {\bibfnamefont {G.}~\bibnamefont
  {Antchev}} \emph {et~al.} (\bibinfo {collaboration} {TOTEM}),\ }\href
  {\doibase 10.1209/0295-5075/101/21003} {\bibfield  {journal} {\bibinfo
  {journal} {Europhys.Lett.}\ }\textbf {\bibinfo {volume} {101}},\ \bibinfo
  {pages} {21003} (\bibinfo {year} {2013}{\natexlab{c}})}\BibitemShut {NoStop}%
\bibitem [{\citenamefont {Abelev}\ \emph {et~al.}(2013)\citenamefont {Abelev}
  \emph {et~al.}}]{Abelev:2012sea}%
  \BibitemOpen
  \bibfield  {author} {\bibinfo {author} {\bibfnamefont {B.}~\bibnamefont
  {Abelev}} \emph {et~al.} (\bibinfo {collaboration} {ALICE Collaboration}),\
  }\href {\doibase 10.1140/epjc/s10052-013-2456-0} {\bibfield  {journal}
  {\bibinfo  {journal} {Eur.Phys.J.}\ }\textbf {\bibinfo {volume} {C73}},\
  \bibinfo {pages} {2456} (\bibinfo {year} {2013})},\ \Eprint
  {http://arxiv.org/abs/1208.4968} {arXiv:1208.4968 [hep-ex]} \BibitemShut
  {NoStop}%
\bibitem [{\citenamefont {Aaij}\ \emph {et~al.}(2015)\citenamefont {Aaij} \emph
  {et~al.}}]{Aaij:2014vfa}%
  \BibitemOpen
  \bibfield  {author} {\bibinfo {author} {\bibfnamefont {R.}~\bibnamefont
  {Aaij}} \emph {et~al.} (\bibinfo {collaboration} {LHCb}),\ }\href {\doibase
  10.1007/JHEP02(2015)129} {\bibfield  {journal} {\bibinfo  {journal} {JHEP}\
  }\textbf {\bibinfo {volume} {1502}},\ \bibinfo {pages} {129} (\bibinfo {year}
  {2015})},\ \Eprint {http://arxiv.org/abs/1412.2500} {arXiv:1412.2500
  [hep-ex]} \BibitemShut {NoStop}%
\bibitem [{\citenamefont {Ostapchenko}(2014)}]{Ostapchenko:2014mna}%
  \BibitemOpen
  \bibfield  {author} {\bibinfo {author} {\bibfnamefont {S.}~\bibnamefont
  {Ostapchenko}},\ }\href {\doibase 10.1103/PhysRevD.89.074009} {\bibfield
  {journal} {\bibinfo  {journal} {Phys.Rev.}\ }\textbf {\bibinfo {volume}
  {D89}},\ \bibinfo {pages} {074009} (\bibinfo {year} {2014})},\ \Eprint
  {http://arxiv.org/abs/1402.5084} {arXiv:1402.5084 [hep-ph]} \BibitemShut
  {NoStop}%
\bibitem [{\citenamefont {Ostapchenko}(2006)}]{Ostapchenko:2005nj}%
  \BibitemOpen
  \bibfield  {author} {\bibinfo {author} {\bibfnamefont {S.}~\bibnamefont
  {Ostapchenko}},\ }\href {\doibase 10.1103/PhysRevD.74.014026} {\bibfield
  {journal} {\bibinfo  {journal} {Phys.Rev.}\ }\textbf {\bibinfo {volume}
  {D74}},\ \bibinfo {pages} {014026} (\bibinfo {year} {2006})},\ \Eprint
  {http://arxiv.org/abs/hep-ph/0505259} {arXiv:hep-ph/0505259 [hep-ph]}
  \BibitemShut {NoStop}%
\bibitem [{\citenamefont {Kohara}\ \emph {et~al.}(2014)\citenamefont {Kohara},
  \citenamefont {Ferreira},\ and\ \citenamefont {Kodama}}]{Kohara:2014cra}%
  \BibitemOpen
  \bibfield  {author} {\bibinfo {author} {\bibfnamefont {A.~K.}\ \bibnamefont
  {Kohara}}, \bibinfo {author} {\bibfnamefont {E.}~\bibnamefont {Ferreira}}, \
  and\ \bibinfo {author} {\bibfnamefont {T.}~\bibnamefont {Kodama}},\ }\href
  {\doibase 10.1088/0954-3899/41/11/115003} {\bibfield  {journal} {\bibinfo
  {journal} {J.Phys.}\ }\textbf {\bibinfo {volume} {G41}},\ \bibinfo {pages}
  {115003} (\bibinfo {year} {2014})},\ \Eprint {http://arxiv.org/abs/1406.5773}
  {arXiv:1406.5773 [hep-ph]} \BibitemShut {NoStop}%
\bibitem [{\citenamefont {Goulianos}(2014)}]{Goulianos:2014hqa}%
  \BibitemOpen
  \bibfield  {author} {\bibinfo {author} {\bibfnamefont {K.}~\bibnamefont
  {Goulianos}},\ }\href {\doibase 10.1051/epjconf/20147100050} {\bibfield
  {journal} {\bibinfo  {journal} {EPJ Web Conf.}\ }\textbf {\bibinfo {volume}
  {71}},\ \bibinfo {pages} {00050} (\bibinfo {year} {2014})}\BibitemShut
  {NoStop}%
\bibitem [{\citenamefont {Abreu}\ \emph {et~al.}(2012)\citenamefont {Abreu}
  \emph {et~al.}}]{Collaboration:2012wt}%
  \BibitemOpen
  \bibfield  {author} {\bibinfo {author} {\bibfnamefont {P.}~\bibnamefont
  {Abreu}} \emph {et~al.} (\bibinfo {collaboration} {Pierre Auger
  Collaboration}),\ }\href {\doibase 10.1103/PhysRevLett.109.062002} {\bibfield
   {journal} {\bibinfo  {journal} {Phys.Rev.Lett.}\ }\textbf {\bibinfo {volume}
  {109}},\ \bibinfo {pages} {062002} (\bibinfo {year} {2012})},\ \Eprint
  {http://arxiv.org/abs/1208.1520} {arXiv:1208.1520 [hep-ex]} \BibitemShut
  {NoStop}%
\bibitem [{\citenamefont {Block}\ and\ \citenamefont
  {Halzen}(2011{\natexlab{b}})}]{Block:2011uy}%
  \BibitemOpen
  \bibfield  {author} {\bibinfo {author} {\bibfnamefont {M.~M.}\ \bibnamefont
  {Block}}\ and\ \bibinfo {author} {\bibfnamefont {F.}~\bibnamefont {Halzen}},\
  }\href {\doibase 10.1103/PhysRevD.83.077901} {\bibfield  {journal} {\bibinfo
  {journal} {Phys.Rev.}\ }\textbf {\bibinfo {volume} {D83}},\ \bibinfo {pages}
  {077901} (\bibinfo {year} {2011}{\natexlab{b}})},\ \Eprint
  {http://arxiv.org/abs/1102.3163} {arXiv:1102.3163 [hep-ph]} \BibitemShut
  {NoStop}%
\bibitem [{\citenamefont {Fagundes}\ \emph
  {et~al.}(2013{\natexlab{b}})\citenamefont {Fagundes}, \citenamefont {Grau},
  \citenamefont {Pacetti}, \citenamefont {Pancheri},\ and\ \citenamefont
  {Srivastava}}]{Fagundes:2013aja}%
  \BibitemOpen
  \bibfield  {author} {\bibinfo {author} {\bibfnamefont {D.~A.}\ \bibnamefont
  {Fagundes}}, \bibinfo {author} {\bibfnamefont {A.}~\bibnamefont {Grau}},
  \bibinfo {author} {\bibfnamefont {S.}~\bibnamefont {Pacetti}}, \bibinfo
  {author} {\bibfnamefont {G.}~\bibnamefont {Pancheri}}, \ and\ \bibinfo
  {author} {\bibfnamefont {Y.~N.}\ \bibnamefont {Srivastava}},\ }\href
  {\doibase 10.1103/PhysRevD.88.094019} {\bibfield  {journal} {\bibinfo
  {journal} {Phys.Rev.}\ }\textbf {\bibinfo {volume} {D88}},\ \bibinfo {pages}
  {094019} (\bibinfo {year} {2013}{\natexlab{b}})},\ \Eprint
  {http://arxiv.org/abs/1306.0452} {arXiv:1306.0452 [hep-ph]} \BibitemShut
  {NoStop}%
\bibitem [{\citenamefont {Phillips}\ and\ \citenamefont
  {Barger}(1973)}]{Phillips:1974vt}%
  \BibitemOpen
  \bibfield  {author} {\bibinfo {author} {\bibfnamefont {R.}~\bibnamefont
  {Phillips}}\ and\ \bibinfo {author} {\bibfnamefont {V.~D.}\ \bibnamefont
  {Barger}},\ }\href {\doibase 10.1016/0370-2693(73)90154-8} {\bibfield
  {journal} {\bibinfo  {journal} {Phys.Lett.}\ }\textbf {\bibinfo {volume}
  {B46}},\ \bibinfo {pages} {412} (\bibinfo {year} {1973})}\BibitemShut
  {NoStop}%
\bibitem [{\citenamefont {Amos}\ \emph {et~al.}(1993)\citenamefont {Amos} \emph
  {et~al.}}]{Amos:1992jw}%
  \BibitemOpen
  \bibfield  {author} {\bibinfo {author} {\bibfnamefont {N.~A.}\ \bibnamefont
  {Amos}} \emph {et~al.} (\bibinfo {collaboration} {E710 Collaboration}),\
  }\href {\doibase 10.1016/0370-2693(93)90707-O} {\bibfield  {journal}
  {\bibinfo  {journal} {Phys.Lett.}\ }\textbf {\bibinfo {volume} {B301}},\
  \bibinfo {pages} {313} (\bibinfo {year} {1993})}\BibitemShut {NoStop}%
\bibitem [{\citenamefont {Ansorge}\ \emph {et~al.}(1986)\citenamefont {Ansorge}
  \emph {et~al.}}]{Ansorge:1986xq}%
  \BibitemOpen
  \bibfield  {author} {\bibinfo {author} {\bibfnamefont {R.}~\bibnamefont
  {Ansorge}} \emph {et~al.} (\bibinfo {collaboration} {UA5 Collaboration}),\
  }\href {\doibase 10.1007/BF01411134} {\bibfield  {journal} {\bibinfo
  {journal} {Z.Phys.}\ }\textbf {\bibinfo {volume} {C33}},\ \bibinfo {pages}
  {175} (\bibinfo {year} {1986})}\BibitemShut {NoStop}%
\bibitem [{\citenamefont {Alner}\ \emph {et~al.}(1987)\citenamefont {Alner}
  \emph {et~al.}}]{Alner:1987wb}%
  \BibitemOpen
  \bibfield  {author} {\bibinfo {author} {\bibfnamefont {G.}~\bibnamefont
  {Alner}} \emph {et~al.} (\bibinfo {collaboration} {UA5 Collaboration}),\
  }\href {\doibase 10.1016/0370-1573(87)90130-X} {\bibfield  {journal}
  {\bibinfo  {journal} {Phys.Rept.}\ }\textbf {\bibinfo {volume} {154}},\
  \bibinfo {pages} {247} (\bibinfo {year} {1987})}\BibitemShut {NoStop}%
\bibitem [{\citenamefont {Bernard}\ \emph {et~al.}(1987)\citenamefont {Bernard}
  \emph {et~al.}}]{Bernard:1986yh}%
  \BibitemOpen
  \bibfield  {author} {\bibinfo {author} {\bibfnamefont {D.}~\bibnamefont
  {Bernard}} \emph {et~al.} (\bibinfo {collaboration} {UA4 Collaboration}),\
  }\href {\doibase 10.1016/0370-2693(87)90285-1} {\bibfield  {journal}
  {\bibinfo  {journal} {Phys.Lett.}\ }\textbf {\bibinfo {volume} {B186}},\
  \bibinfo {pages} {227} (\bibinfo {year} {1987})}\BibitemShut {NoStop}%
\bibitem [{\citenamefont {Armitage}\ \emph {et~al.}(1982)\citenamefont
  {Armitage}, \citenamefont {Benz}, \citenamefont {Bobbink}, \citenamefont
  {Erne}, \citenamefont {Kooijman} \emph {et~al.}}]{Armitage:1981zp}%
  \BibitemOpen
  \bibfield  {author} {\bibinfo {author} {\bibfnamefont {J.}~\bibnamefont
  {Armitage}}, \bibinfo {author} {\bibfnamefont {P.}~\bibnamefont {Benz}},
  \bibinfo {author} {\bibfnamefont {G.}~\bibnamefont {Bobbink}}, \bibinfo
  {author} {\bibfnamefont {F.}~\bibnamefont {Erne}}, \bibinfo {author}
  {\bibfnamefont {P.}~\bibnamefont {Kooijman}},  \emph {et~al.},\ }\href
  {\doibase 10.1016/0550-3213(82)90014-1} {\bibfield  {journal} {\bibinfo
  {journal} {Nucl.Phys.}\ }\textbf {\bibinfo {volume} {B194}},\ \bibinfo
  {pages} {365} (\bibinfo {year} {1982})}\BibitemShut {NoStop}%
\bibitem [{\citenamefont {Affolder}\ \emph {et~al.}(2001)\citenamefont
  {Affolder} \emph {et~al.}}]{Affolder:2001vx}%
  \BibitemOpen
  \bibfield  {author} {\bibinfo {author} {\bibfnamefont {T.}~\bibnamefont
  {Affolder}} \emph {et~al.} (\bibinfo {collaboration} {CDF Collaboration}),\
  }\href {\doibase 10.1103/PhysRevLett.87.141802} {\bibfield  {journal}
  {\bibinfo  {journal} {Phys.Rev.Lett.}\ }\textbf {\bibinfo {volume} {87}},\
  \bibinfo {pages} {141802} (\bibinfo {year} {2001})},\ \Eprint
  {http://arxiv.org/abs/hep-ex/0107070} {arXiv:hep-ex/0107070 [hep-ex]}
  \BibitemShut {NoStop}%
\bibitem [{\citenamefont {Dutta}(2014)}]{Dutta:2014wda}%
  \BibitemOpen
  \bibfield  {author} {\bibinfo {author} {\bibfnamefont {D.}~\bibnamefont
  {Dutta}} (\bibinfo {collaboration} {on behalf of CMS Collaboration}),\
  }\href@noop {} {\  (\bibinfo {year} {2014})},\ \Eprint
  {http://arxiv.org/abs/1412.4977} {arXiv:1412.4977 [hep-ex]} \BibitemShut
  {NoStop}%
\bibitem [{\citenamefont {Engel}\ and\ \citenamefont
  {Ulrich}(2012)}]{Engel:2012pa}%
  \BibitemOpen
  \bibfield  {author} {\bibinfo {author} {\bibfnamefont {R.}~\bibnamefont
  {Engel}}\ and\ \bibinfo {author} {\bibfnamefont {R.}~\bibnamefont {Ulrich}},\
  }\href@noop {} {\bibfield  {journal} {\bibinfo  {journal} {Internal Pierre
  Auger Note}\ }\textbf {\bibinfo {volume} {GAP-2012}} (\bibinfo {year} {March,
  2012})}\BibitemShut {NoStop}%
\bibitem [{\citenamefont {Antchev}\ \emph
  {et~al.}(2013{\natexlab{d}})\citenamefont {Antchev} \emph
  {et~al.}}]{Antchev:2013any}%
  \BibitemOpen
  \bibfield  {author} {\bibinfo {author} {\bibfnamefont {G.}~\bibnamefont
  {Antchev}} \emph {et~al.} (\bibinfo {collaboration} {TOTEM Collaboration}),\
  }\href {\doibase 10.1103/PhysRevLett.111.262001} {\bibfield  {journal}
  {\bibinfo  {journal} {Phys.Rev.Lett.}\ }\textbf {\bibinfo {volume} {111}},\
  \bibinfo {pages} {262001} (\bibinfo {year} {2013}{\natexlab{d}})},\ \Eprint
  {http://arxiv.org/abs/1308.6722} {arXiv:1308.6722 [hep-ex]} \BibitemShut
  {NoStop}%
\bibitem [{\citenamefont {Kopeliovich}(2003)}]{Kopeliovich:2003tz}%
  \BibitemOpen
  \bibfield  {author} {\bibinfo {author} {\bibfnamefont {B.}~\bibnamefont
  {Kopeliovich}},\ }\href {\doibase 10.1103/PhysRevC.68.044906} {\bibfield
  {journal} {\bibinfo  {journal} {Phys.Rev.}\ }\textbf {\bibinfo {volume}
  {C68}},\ \bibinfo {pages} {044906} (\bibinfo {year} {2003})},\ \Eprint
  {http://arxiv.org/abs/nucl-th/0306044} {arXiv:nucl-th/0306044 [nucl-th]}
  \BibitemShut {NoStop}%
\end{thebibliography}%
 \end{document}